\documentclass[12pt,a4]{article}
\usepackage{relsize}
\usepackage{array}
\usepackage{a4wide}
\usepackage[hidelinks]{hyperref}
\usepackage{latexsym}
\usepackage{cite}
\usepackage{comment}
\usepackage{caption}
\usepackage{mathrsfs}
\usepackage{dsfont}
\usepackage{tikz}
\usetikzlibrary{arrows.meta,calc,decorations.markings,math,arrows.meta,shapes.misc, decorations.pathreplacing}
\usepackage{amssymb}
\usepackage{amsmath}
\usepackage{graphicx}
\usepackage{enumitem}
\usepackage{todonotes}
\usepackage{tensor}
\usepackage{comment}
\usepackage{braket}

\usepackage{filecontents}

\newcommand{\tr}{\mathrm{tr}}

\newcommand{\be}{\begin{equation}\label}
\newcommand{\ee}{\end{equation}}
\newcommand{\bea}{\begin{eqnarray}\label}
\newcommand{\eea}{\end{eqnarray}}

\makeatletter
\newcommand*{\textoverline}[1]{$\overline{\hbox{#1}}\m@th$}
\newcommand*\bigcdot{\mathpalette\bigcdot@{.65}}
\newcommand*\bigcdot@[2]{\mathbin{\vcenter{\hbox{\scalebox{#2}{$\m@th#1\bullet$}}}}}
\makeatother

\newcommand{\Npt}{N}

\newcommand{\vex}{\vec{x}}

\makeatletter
\newsavebox{\@brx}
\newcommand{\llangle}[1][]{\savebox{\@brx}{\(\m@th{#1\langle}\)}%
  \mathopen{\copy\@brx\mkern2mu\kern-0.9\wd\@brx\usebox{\@brx}}}
\newcommand{\rrangle}[1][]{\savebox{\@brx}{\(\m@th{#1\rangle}\)}%
  \mathclose{\copy\@brx\mkern2mu\kern-0.9\wd\@brx\usebox{\@brx}}}
\makeatother

\numberwithin{equation}{section} 

\newcommand{\I}{\mathcal{I}}
\newcommand{\Op}{\mathcal{O}}
\newcommand{\sixd}{\text{6d}}
\newcommand{\su}{\frak{su}}

\newcommand{\sOp}{\mathcal{O}}
\newcommand{\fOp}{\Phi}
\newcommand{\fields}{\varphi}

\newcommand{\ward}{Ward-Takahashi}

\begin{document}

\newcommand{\nn}{\nonumber}

 \centerline{\LARGE \bf {\sc Five-Dimensional Path Integrals for} } 
  \vskip 12pt
 \centerline{\LARGE \bf {\sc Six-Dimensional Conformal Field Theories} }

 \vspace{1cm}
  \centerline{
   {\large {\bf  {\sc N.~Lambert,${}^{\,a}$}}\footnote{E-mail address: \href{neil.lambert@kcl.ac.uk}{\tt neil.lambert@kcl.ac.uk}}     \,{\sc A.~Lipstein,$^{\,b}$}\footnote{E-mail address: \href{mailto:arthur.lipstein@durham.ac.uk}{\tt arthur.lipstein@durham.ac.uk}}\, {\sc R.~Mouland${}^{\,c}$}\footnote{E-mail address: \href{rishi.mouland@kcl.ac.uk}{\tt r.mouland@damtp.cam.ac.uk}}   {\sc    and P.~Richmond${}^{\,a}$}}\footnote{E-mail address: \href{mailto:paul.richmond@kcl.ac.uk}{\tt paul.richmond@kcl.ac.uk}}  }  
     
\vspace{1cm}
\centerline{${}^a${\it Department of Mathematics}}
\centerline{{\it King's College London }} 
\centerline{{\it London, WC2R 2LS, UK}} 
  
\vspace{1cm}
\centerline{${}^b${\it Department of Mathematical Sciences}}
\centerline{{\it Durham University}} 
\centerline{{\it  Durham, DH1 3LE, UK}} 

\vspace{1cm}
\centerline{${}^c${\it Department of Applied Mathematics and Theoretical Physics}}
\centerline{{\it University of Cambridge }} 
\centerline{{\it Cambridge, CB3 0WA, UK}}
  
\vspace{12mm}

 
\thispagestyle{empty}

\centerline{\sc Abstract}
\vspace{0.5cm}

In this paper we   derive \ward{} identities from the path integral  of supersymmetric five-dimensional field theories  with an $SU(1,3)$ spacetime symmetry in the presence of instantons. We  explicitly show how    $SU(1,3)$ is enhanced to $SU(1,3)\times U(1)$ where the additional $U(1)$ acts non-perturbatively. Solutions to such \ward{} identities were previously obtained from correlators of six-dimensional Lorentzian conformal field theories but where the instanton number was replaced by the momentum along a null direction. Here we study the reverse procedure whereby we construct correlation functions out of towers of  five-dimensional operators  which satisfy the \ward{} identities of a six-dimensional conformal field theory. This paves the way to computing observables in six dimensions using five-dimensional path integral techniques. We also argue that, once the instanton sector is included into the path integral, the coupling of the five-dimensional Lagrangian must be quantised, leaving no free continuous parameters.

\newpage

\tableofcontents

\section{Introduction}
 
Superconformal field theories in six dimensions play a fundamental role in our understanding of M-theory  and play a central role  in our understanding of quantum field theories in general through compactification to lower-dimensions. On the other hand, their precise formulation remains elusive because conventional Lagrangians with manifest six-dimensional superconformal symmetry do not exist. Despite this difficulty, it is possible to compute many observables in these theories like correlators of protected operators using holography \cite{Bastianelli:1999ab,Bastianelli:1999vm,Eden:2001wg,Arutyunov:2002ff}, conformal bootstrap methods \cite{Heslop:2004du,Beem:2015aoa,Rastelli:2017ymc,Heslop:2017sco,Abl:2019jhh,Alday:2020tgi}, and chiral algebra conjectures \cite{Beem:2014kka,Chester:2018dga}. 

Although it is not possible to write down a Lagrangian with six-dimensional superconformal symmetry, a more useful (and perhaps fundamental) definition of a Lagrangian is one which can be used to compute observables such as correlation functions using a path integral. Indeed a Lagrangian may only manifestly realise    some subgroup of the   symmetries of the full quantum theory, as previously demonstrated by the ABJM theory for M2-branes \cite{Aharony:2008ug}. In a recent series of papers we have constructed  a class of five-dimensional Lagrangians with 12 or 24 superconformal symmetries and a non-Lorentzian $SU(1,3)$ spacetime symmetry \cite{Lambert:2019jwi,Lambert:2019fne,Lambert:2020jjm}, studied their correlators \cite{Lambert:2020zdc} and constructed explicit instanton solutions \cite{Lambert:2021mnu}. In this paper we make a proposal as to how these can be used to provide a path integral construction of correlators of six-dimensional superconformal field theories such as the $(2,0)$ theory associated to M5-branes.

In particular here, using path integral methods,  we derive the conformal \ward{} identities for five-dimensional correlators in the presence of instanton operators. These are local disorder operators that correspond to changing the second Chern number  of the gauge field around an insertion point (see also \cite{Lambert:2014jna,Tachikawa:2015mha,Bergman:2016avc}). The existence of a conserved topological charge given by the instanton number (or more accurately the second Chern number of the gauge fields) leads to an additional $U(1)$ symmetry but one under which all the fields in the Lagrangian are invariant. As we stated above the action has a non-trivial $SU(1,3)$ symmetry; however we will show that this symmetry is broken once we allow for non-trivial topological sectors, corresponding to the insertion of instanton operators. Nevertheless an $SU(1,3)\times U(1)$ symmetry can be restored in the quantum theory, with instanton operators charged under the $U(1)$ factor. Thus the path integral defined using the five-dimensional Lagrangian yields an interacting theory with a manifest and non-trivial $SU(1,3)\times U(1)$ symmetry.

In our previous paper \cite{Lambert:2020zdc} we studied the 
\ward{} identities for the symmetry group $SU(1,3)\times U(1)$ and  showed that 
solutions to them can be obtained from a certain Fourier expansion of the correlators of a  six-dimensional conformal field theory. In this way we showed that the instanton number can be used to encode the Kaluza-Klein momentum along an emergent sixth dimension. A novelty of this reduction is that we use the conformal symmetry of the six-dimensional theory to conformally compactify a null direction. As a result the Fourier expansion reproduces the full correlation functions of non-compact six-dimensional Minkowski space. In particular the $SU(1,3)\times U(1)$ symmetry arises as the subgroup of the conformal group  $SO(2,6)$ that commutes with the Kaluza-Klein momentum operator. 

This therefore leads to a natural proposal about how to go the other way and construct genuine six-dimensional correlators from the five-dimensional theory. However the key question is whether or not the resulting correlators can be identified with those of a six-dimensional Lorentzian conformal field theory. In this paper 
 we discuss some necessary conditions for correlators of the theory to resum to produce six-dimensional correlators invariant under the full $SO(2,6)$. We also argue that once topologically non-trivial sectors of the theory are included the action is no longer single valued on the space of field configurations, unless the inverse coupling constant $k$ is a discrete  orbifold parameter, analogous to that of the ABJM theory.

The rest of this paper is organised as follows. In Section \ref{sec: actsym} we briefly review the Lagrangians described above and their symmetries. In Section \ref{sec: instantons} we allow for more general topologies of the gauge fields through so-called instanton operators and show how the $SU(1,3)$ is broken but then restored by finding a suitable representation of the instanton operators. We also show that once we consider this expanded configuration space we are required to restrict $k$ to discrete values. In Section \ref{sec: 6D} we discuss how to construct correlation functions of a six-dimensional theory and in particular give some necessary conditions for these to satisfy the \ward{} identities of a Lorentzian six-dimensional conformal field theory.  In Section \ref{sec: conclusions} we give our conclusions and discussion on future directions. We also include two appendices.

\section{The Actions and Their Symmetries}\label{sec: actsym}

In this first Section, we review the five-dimensional $\Omega$-deformed gauge theory first introduced in \cite{Lambert:2019jwi} by a reduction of the $(2,0)$ theory, and recast its known spacetime symmetries \cite{Lambert:2019fne} in a language more useful for this paper. There are also $(1,0)$ versions of these actions where the fields further decompose into tensor and hyper multiplets  and the supersymmetries are reduced by a half \cite{Lambert:2020jjm}. The form of the action and symmetries is similar but the hyper multiplet fields are allowed to take values in any representation of the gauge group. In the interests of not introducing additional notation we will not discuss them here since all the results in this paper   extend directly to these theories too as the main tool we exploit is the $SU(1,3)$ symmetry of the action.   

\subsection{Review of Five-dimensional Lagrangian Model}

Our starting point is a non-Abelian but non-Lorentzian gauge theory in five dimensions with arbitrary gauge group. We use the coordinates $(x^-, x^i)$ on $\mathbb{R}^5$, with $i,j,\dots=1,\dots,4$. In addition to its gauge field $A=(A_-, A_i)$, the theory has five scalar fields $X^I$, where $I,J,\dots=6,\dots,10$ and a real 32-component spinor $\Psi$ of $\text{Spin(1,10)}$. Finally, we also have a field $G_{ij}=-G_{ji}$ which is self-dual, $G_{ij}=\frac{1}{2}\epsilon_{ijkl} G_{kl}$. All of the fields $X^I, \Psi$ and $G_{ij}$ transform in the adjoint of the gauge group. 

We choose a $32\times 32$ real representation $\{\Gamma_0,\Gamma_1,\dots, \Gamma_{10}\}$ of the $(1+10)$-dimensional Clifford algebra with signature $(-,+,\dots, +)$, and additionally define the combinations $\Gamma_\pm = (\Gamma_0 \pm \Gamma_5)/\sqrt{2}$ which project onto spinors of definite chirality under $\Gamma_{05}$. The fermion $\Psi$ then satisfies $\Gamma_{012345}\Psi = - \Psi$.

The action of the theory is $S=\int dx^- d^4 x \mathcal{L}$, with
\begin{align}
  \mathcal{L} = \frac{k}{4\pi^2} \text{tr} \bigg\{ & \,\frac{1}{2}F_{-i}F_{-i} - \frac{1}{2}\hat{D}_i X^I \hat{D}_i X^I + \frac{1}{2} \mathcal{F}_{ij} G_{ij}\nn\\
  &\,\,-\frac{i}{2} \bar{\Psi} \Gamma_+ D_- \Psi + \frac{i}{2} \bar{\Psi} \Gamma_i \hat{D}_i \Psi - \frac{1}{2}\bar{\Psi} \Gamma_+ \Gamma^I [X^I, \Psi] \bigg\}\ ,
  \label{eq: Lagrangian}
\end{align}
where $F=dA-iA\wedge A$ is the field strength of $A$, and $D_-, D_i$ are adjoint gauge covariant derivatives for the gauge field $A_-, A_i$, {\it i.e.}\ $D_-=\partial_- - i [A_-,\,\cdot\,]$ and $D_i=\partial_i - i [A_i,\,\cdot\,]$. In terms of these more conventional objects, we have used the corresponding $\Omega$-deformed objects,
\begin{align}
\hat D_i &= D_i - \tfrac12 \Omega_{ij}x^jD_- \ , \nn\\
  \mathcal{F}_{ij}  &= F_{ij}	 - \frac12\Omega_{ik}x^kF_{-j}+\frac12\Omega_{jk}x^kF_{-i}\ ,
\end{align}
where $\Omega_{ij}$ is anti-self-dual and normalised as $\Omega_{ik}\Omega_{jk}=\delta_{ij}$. We also define $\hat{\partial}_i=\partial_- - \tfrac{1}{2}\Omega_{ij} x^j \partial_-$ for later use.
We see  that $G_{ij}$ acts as a Lagrange multiplier, imposing the constraint that $\mathcal{F}_{ij}$ is anti-self-dual, {\it i.e.}\ $\mathcal{F}^+_{ij} =0$, where $\mathcal{F}^+_{ij}=\frac{1}{2}\left(\mathcal{F}_{ij}+\frac{1}{2}\epsilon_{ijkl} \mathcal{F}_{kl}\right)$.

Note that in previous papers we have included   a real variable $R$ with dimensions of length with $\Omega_{ij}\Omega_{jk}=-R^{-2}\delta_{ij}$. However in the current paper we have chosen to absorb $R$ into fields and coordinates. Details of this process, and therefore rules on how to reinstate this parameter, are straightforward and can be found in \cite{Mouland:2021urv}.
As is standard in more conventional Yang-Mills theories, one can perform simple field redefinitions to bring terms quadratic in derivatives to canonical normalisation, and in doing so introduce positive powers of $g$ with  $g^2=4\pi^2/k$ in front of all interaction terms. In this sense, one should think of this $g$ as the coupling of the theory.

Let us comment on the origin of this theory \cite{Lambert:2019jwi} in the case where the gauge group is $U(N_5)$. The AdS/CFT correspondence tells us that the worldvolume theory for a stack of $N_5$ M5-branes is dual to M-theory on an $\text{AdS}_7\times S^4$ background. In analogy with the ABJM construction \cite{Aharony:2008ug} and following the geometric considerations of \cite{Pope:1999xg}, one first considers $\text{AdS}_7$ as a timelike circle fibration $S^1\hookrightarrow \text{AdS}_7\to \tilde{\mathbb{CP}}^3$ over the non-compact complex projective space $\tilde{\mathbb{CP}}^3$. One can then write down a non-Abelian action describing the reduction along the fibre of a stack of M5-branes at fixed $\tilde{\mathbb{CP}}^3$ radius. The geometry suggests such a theory should possess eight real supercharges, and it does. Finally, one takes the embedding radius to infinity, effectively sending the stack of M5-branes to the boundary of $\text{AdS}_7$. This boundary is described by the metric
\begin{align}
  ds^2 = -2\, dx^+\left(dx^- - \frac12 \Omega_{ij}x^idx^j\right) + dx^idx^i \ ,
  \label{eq: metric}
\end{align}
with $x^+\in(-\pi, \pi)$ identified as the coordinate along the fibre along which we have reduced. This metric is of the same conformal class as six-dimensional Minkowski space, and so at the end of the day we have simply performed a conformal compactification of M5-branes on flat space.

Note, as we take the limit to the conformal boundary, certain terms in the action diverge. One is nonetheless able to utilise the technique first described in \cite{Lambert:2019nti} to propose the Lagrangian (\ref{eq: Lagrangian}) to describe the boundary theory.\\

We can then use the geometry of the M5-brane embeddings to predict the symmetries of the theory. Since the metric (\ref{eq: metric}) is conformal to the six-dimensional Minkowski metric, any conformal field theory living on it should realise the full conformal algebra $\frak{so}(2,6)$ as its spacetime symmetries. However, the reduction along the $x^+$ direction breaks $\frak{so}(2,6)$ to the maximal subalgebra $\frak{h}=\frak{su}(1,3)\oplus \frak{u}(1)$ commuting with translations along $x^+$. 

Next, the theory has a manifest $SO(5)$ R-symmetry rotating the scalars $X^I$, corresponding simply in the M5-brane picture to rotations in the directions transverse to the branes.

Finally, the circle reduction breaks only one quarter of the superconformal symmetries, and so we can expect the theory to have 24 real supercharges. This is indeed the case, with 8 realised as rigid supersymmetries, and the remaining 16 as conformal supersymmetries \cite{Lambert:2019jwi}. In the models obtained from $(1,0)$ superconformal field theories  one finds half as many supersymmetries and the R-symmetry is  $SU(2)$ corresponding to a suitable replacement of the $S^5$ factor. 

\subsection{Spacetime Symmetry Algebra}

Let us now review the spacetime symmetry structure of the theory in more detail\footnote{A more comprehensive review can be found in \cite{Mouland:2021urv}.}. The subalgebra $\frak{h}\subset \frak{so}(2,6)$ is spanned by the generators $\mathcal{B}=\{P_-, P_i, B, C^\alpha, T, M_{i+}, K_+\}$ along with central element $P_+$, which is simply the generator of translations along the $x^+$ direction along which we have reduced. The other generators have the following action on the five-dimensional coordinates $(x^-, x^i)$
\begin{itemize}\setlength\itemsep{0.2em}
  \item $\{P_-,P_i\}$ are five translations, which form a non-Abelian subalgebra, 
  \item $\{B,C^\alpha\}$, $\alpha=1,2,3$, form a $\frak{u}(1)\oplus \frak{su}(2)$ subalgebra of four rotations in the $x^i$ directions,
  \item $T$ is a Lifshitz scaling, under which $x^-$ scales twice as quickly as $x^i$,
  \item $\{M_{i+},K_+\}$ are `special' transformations, which play much the same role as special conformal transformations in the conformal algebra.
\end{itemize}
A subset of the commutation relations of the algebra is
\begin{align}
[M_{i+},P_j] \ &= \ -\delta_{ij} P_+ - \tfrac{1}{2}\Omega_{ij} T - 2 \delta_{ij} B + \Omega_{ik}\eta^\alpha_{jk} C^\alpha    \, ,	&	[T,P_-] \ 	&= \ -2P_-					\, ,\nn\\
  [T,K_+] \	 &= \ 2K_+ \, ,					&		[P_-,P_i] 				\ &= \ 0		  				\, ,\nn\\
  [K_+,P_-] \ &= \ -2T	\, ,					&		[P_-,M_{i+}] 		\ &= \ P_i   			\, ,\nn\\
[M_{i+},M_{j+}] \ &= \ -\tfrac{1}{2} \Omega_{ij} K_+ \, , &										 		[K_+,P_i] 				\ &= \ -2M_{i+} 		\, ,\nn\\
  [T,P_i] 	\ &= \ -P_i \, ,		&		[K_+,M_{i+}] 			\ &= \  0  						\, ,\nn\\
  [T,M_{i+}] \ &= \ M_{i+} \, ,	&		[P_i,P_j] 		\ &= \ -\Omega_{ij} P_-  						\, .
\label{eq: extended su(1,3) algebra}
\end{align}
The rotations $B,C^\alpha$ form an $\frak{u}(1)\oplus \frak{su}(2)$ subalgebra;
\begin{align}
  [B,C^\alpha]	\ = \ 0\, ,\qquad [C^\alpha, C^\beta] \ = \ -\varepsilon^{\alpha\beta\gamma}C^\gamma \, .
  \label{eq: rotation subalgebra}
\end{align}
In particular these generate all rotations in the four-dimensional plane that leave $\Omega_{ij}$ invariant. 
The remaining brackets are neatly summarised by noting that the `scalar' generators $\mathcal{S}=P_-,T, K_+$ are inert under the rotation subgroup, {\it i.e.}\ $[\mathcal{S},B]=[\mathcal{S},C^\alpha]=0$, while the `one-form' generators $W_i=P_i,M_{i+}$ transform as
\begin{align}
  	[W_i,B] 	\ &= \ -\tfrac{1}{2} \,\Omega_{ij} W_j 	\, ,	\qquad [W_i,C^\alpha] \	= \ \tfrac{1}{2}\eta^\alpha_{ij} W_j \, .
  	\label{eq: rotation subalgebra action}
\end{align}
If we for a moment exclude the central element $P_+$, then the elements of $\mathcal{B}$ form a (somewhat unconventional) basis for $\frak{su}(1,3)$. In fact, the centrally extended algebra $\frak{h}$ can be realised as simply $\frak{h}=\frak{su}(1,3)\oplus \frak{u}(1)$, with basis $\{P_-, P_i, \tilde{B}, C^\alpha, T, M_{i+}, K_+\}$ for the $\frak{su}(1,3)$ factor and $P_+$ for the $\frak{u}(1)$ factor. Here, we have $\tilde{B}=B+\tfrac{1}{2}P_+$. However, it will be more convenient for geometric reasons to continue to use $B$ rather than $\tilde{B}$, and thus refrain from making this direct sum decomposition of $\frak{h}$ manifest.

\subsection{Realisation on Coordinates and Fields}

Let us now investigate how these symmetries are realised by the Lagrangian (\ref{eq: Lagrangian}). There is a little nuance here regarding what we should expect. We interpret $S$ as describing $N$ M5-branes reduced along the direction $x^+\in(-\pi , \pi )$. Let us first suppose, as in a standard Kaluza-Klein reduction, that in doing this reduction we have truncated the spectrum of the theory maximally; in other words, the theory $S$ describes only the zero modes on the $x^+$ interval. We know then that such modes will fall into representation of $\frak{h}$ in which $P_+$ is represented trivially ({\it i.e.}\ it annihilates everything); in other words, representations of $\frak{su}(1,3)$. Thus, we expect $S$ to admit an $\frak{su}(1,3)$ spacetime symmetry.

Conversely, just as five-dimensional maximal super-Yang-Mills is conjectured to in fact describe \textit{all} modes of a spatial compactification of M5-branes through the inclusion of local operators with non-zero instanton charge \cite{Lambert:2010iw,Douglas:2010iu}, we also propose that our action $S$ should describe all modes of the $x^+$ conformal compactification. Modes with non-zero charge under $P_+$ are expected to be realised only when the configuration space is extended to allow for isolated singular points, around which one measures non-zero instanton number.

What we will show first is that if we disallow such configurations, then the theory does indeed admit an $\frak{su}(1,3)$ spacetime symmetry. It will already be clear however at this point that something goes wrong when the configuration space is extended. We will indeed show below that in this case we precisely recover modes with non-trivial charge under $P_+$, and thus the spacetime symmetry algebra is extended to $\frak{h}$.

So let us first describe the $\frak{su}(1,3)$ spacetime symmetry of the theory, as first discussed in \cite{Lambert:2019fne}, which is valid when the gauge field is regular throughout $\mathbb{R}^5$. Our first step is to define some action of $\frak{su}(1,3)$ on coordinates and fields. As a spacetime symmetry, $\frak{su}(1,3)$ admits a representation in terms of vector fields on $\mathbb{R}^5$. Given some $G\in\frak{su}(1,3)$, we have corresponding vectors fields $G_\partial$, with
\begin{align}
	\left( P_- \right)_\partial \ 	&= \ \partial_-  \, ,  \nn\\
	\left( P_i \right)_\partial \ 	&= \ \tfrac{1}{2}\Omega_{ij} x^j \partial_- + \partial_i	\, , \nn\\
	\left( B \right)_\partial 		\ &= \ -\tfrac{1}{2}\,\Omega_{ij}x^i\partial_j 		\, , \nn\\
	\left( C^\alpha \right)_\partial 	\ &= \ \tfrac{1}{2}\eta^\alpha_{ij}x^i\partial_j 								\, , \nn\\
	\left( T \right)_\partial 		\ &= \ 2x^- \partial_- + x^i \partial_i					\, , \nn\\
	\left( M_{i+} \right)_\partial 	\ &= \ \left( \tfrac{1}{2}\Omega_{ij} x^- x^j - \tfrac{1}{8} x^j x^j x^i \right)\partial_- + x^- \partial_i  +\tfrac{1}{4}( 2\Omega_{ik}x^k x^j + 2\Omega_{jk}x^k x^i - \Omega_{ij}x^k x^k )\partial_j	\, , \nn\\
	\left( K_{+} \right)_\partial 	 \ &= \ ( 2 ( x^- )^2 - \tfrac{1}{8}  ( x^i x^i )^2 )\partial_- +( \tfrac{1}{2} \Omega_{ij} x^j x^k x^k + 2 x^- x^i )\partial_i	\, .
	\label{eq: 5d algebra vector field rep}
\end{align}
Let us set up our conventions for $SU(1,3)$ transformations. Given any $g=e^{\epsilon G}\in SU(1,3)$, and any point $x\in\mathbb{R}^5$, we can denote by $xg\in\mathbb{R}^5$ the point sitting at a finite distance $\epsilon$ along the integral curve of $G_\partial$ starting at $x$. Then, we have for $g_1,g_2\in SU(1,3)$, $x(g_1 g_2)=(xg_1)g_2$, and so $SU(1,3)$ admits a natural right action on $\mathbb{R}^5$. In this way, we can consider $SU(1,3)$ orbits on our spacetime. For infinitesimal $G$, the leading order term in $xg$ can be read off from (\ref{eq: 5d algebra vector field rep}), while the finite form of $xg$ for $g$ generated by each of the basis generators can be found in \cite{Mouland:2021urv}.

Next, we consider how some generic field in the theory, which we denote by $\fields$,  transforms under $SU(1,3)$. Under an active $SU(1,3)$ transformation $g$, we have
\begin{align}
  x		&\quad\longrightarrow\quad x		\nn\\
  \fields(x) &\quad\longrightarrow\quad \fields'(x) = g \fields(x) :=  \mathcal{R}_g(xg^{-1})\fields(xg^{-1})\ ,
  \label{eq: general field transformation}
\end{align}
where $\mathcal{R}_g$ is some (generically spacetime-dependent) matrix acting on any indices of $\fields$, and satisfying $\mathcal{R}_{g_2}(xg_1)\mathcal{R}_{g_1}(x)=\mathcal{R}_{g_1 g_2}(x)$. Taking $g$ then to act only on fields, so that for instance $g(\partial_i \fields(x))=\partial_i \left( g\fields (x) \right)$, we have that $(g_1 g_2)\fields(x) = g_1\left( g_2\fields(x) \right)$.

For $G$ infinitesimal, we can write to leading order\footnote{In our conventions, $\delta_G$ acts on fields only, so that for instance $\delta_G\left( x^i\partial_i \fields(x) \right)=x^i\partial_i\left( \delta_G \fields(x) \right)$.} $g\fields(x)=\fields(x) + \delta_G\fields(x)$, where $\delta_G\fields(x) = -G_\partial \fields(x) - r_G(x)\fields(x)$. $r_G(x)$ is a matrix acting on any indices of $\fields(x)$, and satisfying $[r_{G_1},r_{G_2}]+\left(G_1\right)_\partial r_{G_2} - \left(G_1\right)_\partial r_{G_1}=r_{[G_1,G_2]}$ for any $G_1, G_2\in\frak{su}(1,3)$. Then, the variations $\delta_G$ form a representation of $\frak{su}(1,3)$, {\it i.e.}\ $[\delta_{G_1},\delta_{G_2}]=\delta_{[G_1, G_2]}$.

The general form of the $r_G(x)$ can be deduced by defining a notion of primaries and descendants of $\frak{su}(1,3)$ \cite{Lambert:2020zdc}. In particular,  primaries are annihilated at the origin by the special transformations $M_{i+}, K_+$, with descendants generated by the action of $P_-, P_i$. A primary operator is entirely captured by a Lifshitz scaling dimension $\Delta$ and representations $r[B], r[C^\alpha]$ under the rotation subalgebra. Explicitly, then, such a primary transforms under $\frak{su}(1,3)$ as
\begin{align}
  	\delta_{P_-}\fields(x)		\ &=	 \ -\left( P_- \right)_\partial \fields(x)	\ ,							\nn\\
  	\delta_{P_i}\fields(x)		\ &= \ -\left( P_i \right)_\partial \fields(x)	\ ,							\nn\\
  	\delta_{B}\fields(x)		\ &=	 \ -\left( B \right)_\partial \fields(x) - r_\fields[B]   \fields(x)\ ,	\nn\\
  	\delta_{C^\alpha}\fields(x)		\ &=	 \ -\left( C^\alpha \right)_\partial \fields(x) - r_\fields[C^\alpha] \fields(x)	\ ,\nn\\
  	\delta_{T}\fields(x)		\ &= \	-\left( T \right)_\partial \fields(x) -  \Delta \fields(x)	\ ,\nn\\
  	\delta_{M_{i+}}\fields(x)	\ &= \	-\left( M_{i+} \right)_\partial \fields(x) - \left(\tfrac{1}{2}\Delta \Omega_{ij} x^j  +2x^i r_\fields[B] - \Omega_{ik} \eta^\alpha_{jk} x^j r_\fields[C^\alpha]\right) \fields(x)			\ ,\nn\\
  	\delta_{K_+}\fields(x)			\ &=	 \ -\left( K_+ \right)_\partial \fields(x)  - \left(2\Delta\, x^-  +2x^i x^i r_\fields[B] - x^i x^j \Omega_{ik}\eta^\alpha_{jk} r_\fields[C^\alpha]   \right) \fields(x) \ .
  	\label{eq: primary of su(1,3)}
\end{align}
The gauge field $A$, scalars $X^I$ and fermions $\Psi$ do indeed fall into representations of $\frak{su}(1,3)$ and thus transform as in (\ref{eq: general field transformation}) for some non-trivial variations $\delta_G$. Hence, these fields can be reorganised (albeit somewhat non-trivially) to be written in terms of such $\frak{su}(1,3)$ primaries \cite{Mouland:2021urv}.
Full details of the infinitesimal transformations of fields can be found in Appendix \ref{app: field variations}, while their finite transformations are also known \cite{Mouland:2021urv}, but will not be needed here.

The Lagrange multiplier $G_{ij}$ is a little different. The variation $\delta_G G_{ij}$ depends not only on $G_{ij}$ but also the field strength $F$ of the gauge field, at least for $G=M_{i+}, K_+$. Thus, one should really regard $(A,G_{ij})$ sitting in a single representation. Further, the algebra of variations $\delta_G$ only closes on $G_{ij}$ on-shell; more specifically, it closes only on the constraint surface $\mathcal{F}_{ij}=-\star\mathcal{F}_{ij}$.

Finally, it is notationally convenient to introduce a trivial variation $\delta_{P_+}$, acting as $\delta_{P_+}X^I=0$, $\delta_{P_+}\Psi=0$, $\delta_{P_+}A=0$ and $\delta_{P_+}G_{ij}=0$. 
Then, the $\{\delta_G\}_{G\in \mathcal{B} \cup \{P_+\}}$ generate a representation of $\frak{h}$ with $P_+$ trivially represented, and we have $[\delta_{G_1},\delta_{G_2}]=\delta_{[G_1,G_2]}$ for all $G_1,G_2\in\frak{h}$, with brackets as given in (\ref{eq: extended su(1,3) algebra})--(\ref{eq: rotation subalgebra action}).

\subsection{Variation of the Lagrangian}

We have shown that the full field content of the theory falls into representations of $\frak{h}$ (at least on the constraint surface, in the case of $G_{ij}$) under the variations $\delta_G$. We reiterate, these are indeed representations of $\frak{su}(1,3)\subset \frak{h}$, as $P_+$ is trivially represented: $\delta_{P_+}\fields=0$ on all fields $\fields=A, X^I, \Psi, G_{ij}$. Further, the Lagrangian (\ref{eq: Lagrangian}) transforms in a representation\footnote{This is slightly non-trivial, since the algebra does \textit{not} close off-shell on $G_{ij}$. However, the corresponding anomalous extension to the symmetry algebra (\ref{eq: Gij algebra extension}) is parameterised by an additional variation $\bar{\delta}_{ij}$, which explicitly annihilates the Lagrangian. Thus, the algebra does close on $\mathcal{L}$ off-shell.} of $\frak{su}(1,3)\subset\frak{h}$.

So let us state the variation of the Lagrangian $\mathcal{L}$. In addition to the trivial $\delta_{P_+}\mathcal{L}=0$, for $G\in\{P_-, P_i, B, C^\alpha, T\}$ we find
\begin{align}
  -\delta_{P_-}\mathcal{L}	&= \partial_- \mathcal{L}		\ ,\nn\\
  -\delta_{P_i}\mathcal{L}	&= \partial_- \left( \tfrac{1}{2}\Omega_{ij} x^j \mathcal{L} \right) + \partial_i \mathcal{L} 	\ ,\nn\\
  -\delta_{B}\mathcal{L}		&= \partial_i \left( \tfrac{1}{2} \Omega_{ij} x^j\mathcal{L} \right) 			\ ,\nn\\
  -\delta_{C^\alpha}\mathcal{L}	&= \partial_i \left( -\tfrac{1}{2} \eta^I_{ij} x^j\mathcal{L} \right) 				\ ,\nn\\
  -\delta_{T}\mathcal{L}		&= \partial_- \left( 2x^-\mathcal{L} \right) + \partial_i \left( x^i\mathcal{L} \right) 	\ ,
\end{align}
and hence with suitable boundary conditions on the 4-sphere $S^4_\infty$ at infinity, we have $\delta_G S=0$. More care must be taken, however, in the case of $G\in\{M_{i+}, K_+\}$. We find\footnote{Here and throughout, we take $\star$ to denote the Hodge star with respect to the Euclidean metric on $\mathbb{R}^5$, which satisfies $\star^2=1$ on all forms, and $\left( \star d \star \omega \right)_{\alpha_1\dots \alpha_{p-1}}=\partial^\beta \omega_{\alpha_1\dots \alpha_{p-1} \beta}$ for generic $p$-form $\omega$.}
{\allowdisplaybreaks
\begin{align}
  -\delta_{M_{i+}} \mathcal{L} &= \star \left( dx^i \wedge \left( \frac{k}{8\pi^2}\tr\left( F\wedge F \right) \right) \right)		\nn\\*
  &\qquad+   \partial_- \left[ \left( \frac{1}{2}\Omega_{ij}x^- x^j - \frac{1}{8}x^j x^j x^i \right)\mathcal{L} -\frac{k}{16\pi^2} x^i \,\text{tr}\big( X^I X^I \big) \right] 	 				\nn\\*
  &\qquad+ \partial_j \left[ \frac{1}{4}\left( 2\Omega_{ik} x^k x^j + 2\Omega_{jk} x^k x^i - \Omega_{ij} x^k x^k +4x^- \delta_{ij} \right)\mathcal{L} -\frac{k}{8\pi^2} \Omega_{ij} \,\text{tr}\big( X^I X^I \big) \right]	\ ,\nn\\
  -\delta_{K_+} \mathcal{L} &= \star \left( d\left( x^i x^i \right) \wedge \left( \frac{k}{8\pi^2}\tr\left( F\wedge F \right) \right) \right)		\nn\\*
  &\qquad+   \partial_- \left[ \left( 2\left( x^- \right)^2-\frac{1}{8}(x^i x^i)^2 \right)\mathcal{L} -\frac{k}{8\pi^2} x^ix^i \,\text{tr}\big( X^I X^I \big) \right] 	 				\nn\\*
  &\qquad+ \partial_i \left[ \left( \frac{1}{2}\Omega_{ij} x^j x^k x^k+2 x^- x^i \right)\mathcal{L} +\frac{k}{4\pi^2} \Omega_{ij}x^j \,\text{tr}\big( X^I X^I \big) \right]\ .
\end{align}
}%
If we require that the gauge field $A$ is globally defined and regular everywhere, then we can write
\begin{align}
  dx^i \wedge \left( \frac{k}{8\pi^2}\tr\left( F\wedge F \right) \right) &= d\left( \frac{k}{8\pi^2} x^i \tr\left( F\wedge F \right) \right)	\ ,\nn\\
  d\left( x^i x^i \right) \wedge \left( \frac{k}{8\pi^2}\tr\left( F\wedge F \right) \right) &= d\left( \frac{k}{8\pi^2} x^i x^i \tr\left( F\wedge F \right) \right)\ ,
\end{align}
and hence, in both cases, $\delta_G\mathcal{L}$ is a total derivative, and for suitable boundary conditions on $S^4_\infty$ we have $\delta_G S=0$.

\section{Instantons}\label{sec: instantons}

We have now seen that the theory described by Lagrangian (\ref{eq: Lagrangian}) does indeed possess an $\frak{su}(1,3)$ spacetime symmetry when the gauge field $A$ is regular throughout $\mathbb{R}^5$. It would therefore be reasonable to propose that the theory describes only the zero modes of the compactification on $x^+\in(-\pi , \pi )$, since it admits a symmetry under $\frak{h}$ in which nothing is charged under $P_+$. To move beyond this, we now instead consider a broader configuration space for the theory. 

\subsection{Instantons and Classical Symmetry Breaking}\label{subsec: classical instantons}

Our task now is to broaden the class of spaces we allow our theory to live on, in an effort to introduce non-trivial topological sectors of the configuration space. Let us now and for the remainder of this paper specialise to gauge group $G=SU(N_c)$.

It is clear that all principal $SU(N_c)$ bundles $P\to \mathbb{R}^5$ are trivialisable. Consider instead however removing a set of points $\{x_a\}_{a=1}^M$ and considering  principal bundles over $M_5=\mathbb{R}^5\setminus\{x_a\}_{a=1}^M$. Such bundles are then characterised by the integral of the second Chern class over small 4-spheres surrounding each of the $x_a$, which are quantised as
\begin{align}
  n_a = \frac{1}{8\pi^2} \int_{S^4_a} \text{tr}\left( F\wedge F \right) \in\mathbb{Z}\ ,
  \label{eq: flux quantisation}
\end{align}
with $S^4_a$ denoting a small 4-sphere surrounding the puncture at $x_a$. We then call each pair $(x_a, n_a)$ an \textit{instanton insertion}, with $x_a\in\mathbb{R}^5$ the instanton insertions's \textit{position}, and $n_a\in\mathbb{Z}$ its \textit{charge}. We could also in principle consider allowing for non-zero instanton number on $S^4_\infty$, but we instead consider only configurations with
\begin{align}
  \frac{1}{8\pi^2}\int_{S^4_\infty} \tr \left( F\wedge F \right) = 0\ .
  \label{eq: vanishing flux at infinity}
\end{align}
Since the finite $SU(1,3)$ transformations generated by $M_{i+}$ and $K_+$ move the point at infinity \cite{Mouland:2021urv}, this is chosen as a convenience, rather than a restriction.

Note then that since $d\,\tr\left( F\wedge F \right)=0$ throughout $M_5$, we have
\begin{align}
  0 = \frac{1}{8\pi^2}\int_{S^4_\infty} \tr \left( F\wedge F \right) = \sum_{a=1}^M \frac{1}{8\pi^2} \int_{S^4_a} \text{tr}\left( F\wedge F \right) = \sum_{a=1}^M n_a\ .
  \label{eq: sum of charges is zero}
\end{align}
Thus, the data of the bundle is contained within the set of instanton insertions $\{(x_a, n_a)\}_{a=1}^M$, with the $x_a$ distinct and the $n_a$ summing to zero. Necessarily, $M_5$ must now be covered in a number of patches, on each of which $A$ is defined. One can however consider a limit of such an open cover, such that $A$ is now globally defined and regular except along 1-dimensional strings where it is singular. These strings, which are analogous to the Dirac string, extend between the insertions $x_a$. Then, the integral of the Chern-Simons 3-form on any $S^3$ through which such a string is threaded is quantised, ensuring that (\ref{eq: flux quantisation}) is satisfied. Gauge field configurations with precisely this form were   found in \cite{Lambert:2021mnu}, but we will not need their details here.

We are now able to extend our field content back to the whole of $\mathbb{R}^5$, so long as we allow for particular singular behaviour of the field strength $F$. We in particular have
\begin{align}
  d\left( \frac{1}{8\pi^2}\tr\left( F\wedge F \right) \right) = d^5 x\,\sum_{a=1}^M n_a \delta^{(5)}(x-x_a)\ .
\end{align}
Such configurations with maximal symmetry about the points $x_a$ will behave as
\begin{align}
  \frac{1}{8\pi^2}\tr\left( F\wedge F \right) \sim  -\frac{n_a}{6\pi^2}\star d\left( \frac{1}{|x-x_a|^3} \right)\ ,
  \label{eq: sph symmetric instanton}
\end{align}
as we approach $|x-x_a|\to 0$, where here $|x|^2=(x^-)^2+x^i x^i$. However, we more generally only expect the pullback to the $S^4$ surrounding $x_a$ to behave as
  \begin{align}
  \left.\frac{1}{8\pi^2}\tr\left( F\wedge F \right)\,\right|_{S^4} \sim n_a\Omega_4\ ,
\end{align}
as we approach $|x-x_a|\to 0$, where $\Omega_4$ encodes angular dependence, and satisfies $\int_{S^4}\Omega_4=1$. Thus, the components of $\left.\frac{1}{8\pi^2}\tr\left( F\wedge F \right)\,\right|_{S^4}$ in Cartesian coordinates on $\mathbb{R}^5$ go as $|x-x_a|^{-4}$ as we approach $|x-x_a|\to 0$.

Explicit examples of such configurations on $S^4$ can be constructed by suitable stereographic projection from corresponding configurations on $\mathbb{R}^4$. The minimal such construction \cite{Bergman:2016avc}, in which the $SU(2)$ BPST instanton of size $\rho$ is mapped to $S^4$, corresponds to $n_a=\pm 1$, with $\rho=1$ producing the spherically symmetric result (\ref{eq: sph symmetric instanton}). More generally, one can in principle relate\footnote{Note that generically some moduli that are physical on $\mathbb{R}^4$ can become gauge redundancies on $S^4$ - for instance, the three gauge orientation moduli of the single $SU(2)$ instanton.} any $SU(N_c)$ $n$-instanton configuration on $\mathbb{R}^4$, parameterised by $4nN_c$ moduli and captured by the ADHM construction \cite{Atiyah:1978ri}, to a corresponding configuration on $S^4$ by stereographic projection. While we will not require any of the finer details of such constructions, it is important to emphasise that just specifying instanton insertions $\{(x_a,n_a)\}$ does \textit{not} fix the boundary behaviour of the gauge field $A$ in a neighbourhood of the points $\{x_a\}$, but rather specifies that such behaviour belongs to a particular continuous family of instanton profiles.

Further, note that configurations defined over $\mathbb{R}^5$ which feature an arbitrary number of instanton insertions at points $x_a$, as well as vanishing flux on $S^4_\infty$ as in (\ref{eq: vanishing flux at infinity}), were found in \cite{Lambert:2021mnu}. Such configurations additionally satisfy the constraint $\mathcal{F}_{ij}+\frac{1}{2}\varepsilon_{ijkl}\mathcal{F}_{kl} =0$ imposed by $G_{ij}$.

So, we now take the configuration space of our theory to be extended to a disjoint union of subspaces, on each of which we specify instanton insertions $\{(x_a,n_a)\}$. Note, for the sake of later notational convenience, we allow for any of the $n_a$ to be zero, in which case $F$ can be smoothly extended to $x_a$. 

It is crucial to note that $SU(1,3)$ still admits an action on this extended configuration space. In particular, the form of $gA$ ensures that
\begin{align}
  d\left( \frac{1}{8\pi^2}\tr\left( F[A]\wedge F[A] \right) \right) = d^5 x\,\sum_{a=1}^M n_a \delta^{(5)}(x-x_a) \nn\\ 
  \implies \quad  d\left( \frac{1}{8\pi^2}\tr\left( F[gA]\wedge F[gA] \right) \right) &= d^5 ( x g^{-1} )\,\sum_{a=1}^M n_a \delta^{(5)}(xg^{-1}-x_a)		\nn\\
  &= d^5 x\sum_{a=1}^M n_a \delta^{(5)}(x-x_ag)\ ,
  \label{eq: moving of instanton insertions}
\end{align}
and hence if $A$ has instanton insertions $\{(x_a, n_a)\}_{a=1}^M$, the transformed field $gA$ has instanton insertions $\{(x_ag, n_a)\}_{a=1}^M$.\\

Let us now return to the $\su(1,3)$ variation of the Lagrangian. We find now that in the presence of instanton insertions, the variation of $\mathcal{L}$ under $M_{i+}, K_+$ is no longer a total derivative, and the action is no longer invariant. We find
\begin{align}
  	\delta_{M_{i+}}\mathcal{L} &= k\sum_{a=1}^M n_a x_a^i \delta^{(5)}(x-x_a) + \star\, d \left( \dots \right)		\ ,\nn\\
  	\delta_{K_+}\mathcal{L} &= k\sum_{a=1}^M n_a x_a^i x_a^i \delta^{(5)}(x-x_a) + \star\, d \left( \dots \right)		\ ,
\end{align}
and hence, for suitable boundary conditions on $S^4_\infty$, we have
\begin{align}
  	\delta_{M_{i+}}S &= k\sum_{a=1}^M n_a x_a^i		\ ,\nn\\
  	\delta_{K_+}S &= k\sum_{a=1}^M n_a x_a^i x_a^i		\ .
  	\label{eq: M and K action variation}
\end{align}
Thus, we find that the classical action is no longer invariant under $SU(1,3)$. However, we note that the variation of the action is \textit{local} to the punctures $\{x_a\}$. It is precisely this fact that allows for a recasting of the classical non-invariance of $S$ as a symmetry \textit{deformation} in the quantum theory.

However, before exploring this we finally note the transformation of the action under the finite transformations generated by $M_{i+}$ and $K_+$, which are found by exponentiating the infinitesimal results (\ref{eq: M and K action variation}).

Again let $\fields=A, X^I, \Psi, G_{ij}$ be shorthand for the set of fields of the theory, and suppose that the gauge field $A$ has insertions $\{(x_a, n_a)\}_{a=1}^M$. Then, we find
\begin{align}
  S[e^{\epsilon^i M_{i+}}\fields] = S[\fields] - i k \sum_{a=1}^M n_a \log \left( \frac{\overline{\mathcal{M}_\epsilon(x_a)}}{\mathcal{M}_\epsilon(x_a)} \right)\ ,
\end{align}
where given some 4-vector $\alpha^i$, we define
\begin{align}
  \mathcal{M}_\alpha(x) &= \left( 1-\tfrac{1}{2}\Omega_{ij}\alpha^i x^j+\tfrac{1}{16}\alpha^i \alpha^i x^j x^j \right) - \frac{i}{4}\left( \alpha^i \alpha^i x^- + 2\alpha^i x^i \right)\nn\\
  &= 1+z(x,(0,\alpha^i))-z(x,0) - z(0,(0,\alpha^i))-\frac{i}{4}\alpha^i \alpha^i z(x,0)\ ,
  \label{eq: curly M definition}
\end{align}
and we have the complex distance
\begin{align}
  z(x_1,x_2)=\left( x_1^--x_2^-+\frac{1}{2}\Omega_{ij} x_1^i x_2^j \right) + \frac{i}{4}(x_1^i-x_2^i)(x_1^i-x_2^i) = -\bar{z}(x_2,x_1)\ .\label{zis}
\end{align}
Equivalently, we can write
\begin{align}
  \exp\left( {iS[e^{\epsilon^i M_{i+}}\fields]} \right) = e^{iS[\fields]} \prod_{a=1}^M \left( \frac{\overline{\mathcal{M}_\epsilon(x_a)}}{\mathcal{M}_\epsilon(x_a)} \right)^{kn_a}\ .
  \label{eq: finite action variation M}
\end{align}
Similarly, we find
\begin{align}
  S[e^{\epsilon K_+}\fields] = S[\fields] - ik\sum_{a=1}^M n_a\log \left( \frac{1-2\epsilon \bar{z}(x_a,0)}{1-2\epsilon z(x_a,0)} \right)\ ,
  \label{eq: finite action variation K just action}
\end{align}
or equivalently,
\begin{align}
  \exp\left( {iS[e^{\epsilon K_+}\fields]}  \right)= e^{iS[\fields]} \prod_{a=1}^M \left( \frac{1-2\epsilon \bar{z}(x_a,0)}{1-2\epsilon z(x_a,0)} \right)^{kn_a}\ .
  \label{eq: finite action variation K}
\end{align}
Note that the multiplicative factors appearing on the right hand side of (\ref{eq: finite action variation M}) and (\ref{eq: finite action variation K}) generically have branch points. This suggests that there may exist closed loops in configurations space, around which $e^{iS}$ picks up a non-trivial phase. The existence of such loops would thus signal a failure of single-valuedness of $e^{iS}$ as a functional on configuration space. This will be explored in Section \ref{sec: k}.

\subsection{Quantum Recovery}

We now consider the fate of our $\frak{su}(1,3)$ symmetry in the corresponding quantum theory. Despite the non-invariance of the action, we find a set of \ward{} identities satisfied by all correlation functions of the theory.

Such identities are of the usual form, in particular involving the divergence of some vector current; the Noether current for the respective symmetry. The derivation of such \textit{local} \ward{} identities and corresponding currents is left until Section \ref{subsec: local WTIs}. We first derive the corresponding \textit{global} identities---also obtainable by integrating their local counterparts over $\mathbb{R}^5$---directly, so as to elucidate the quantum recovery of the theory's symmetries most straightforwardly.\\

First suppose we forbid instanton insertions, and define the configuration space of the theory to have globally regular field strength $F$.  We can then formally define correlation functions of operators $\fOp^{(1)},\dots,\fOp^{(N)}$ by the path integral
\begin{align}
  \left\langle \fOp^{(1)}(x_1)\dots \fOp^{(N)}(x_N) \right\rangle = \int D\fields\, \fOp^{(1)}(x_1)\dots \fOp^{(N)}(x_N) e^{iS[\fields]}\ ,
  \label{eq: basic path integral}
\end{align}
where, as above, we use $\fields$ to denote the fields $X^I, A, \Psi, G_{ij}$ of the theory, and the $\fOp^{(a)}$ are generically composite functions of $\fields$ and their derivatives. The partition function is $\mathcal{Z}=\langle\mathds{1}\rangle$.

Symmetries are then realised by \ward{} identities for correlations functions. Under some $SU(1,3)$ transformation $g$, we have transformed fields $\fields'=g\fields$. Making use of the fact that $S[\fields']=S[\fields]$, and assuming $D\fields' = D\fields$, we have
\begin{align}
  \langle \fOp^{(1)'}(x_1)\dots \fOp^{(N)'}(x_N) \rangle &= \int D\fields\, \fOp^{(1)'}(x_1)\dots \fOp^{(N)'}(x_N) e^{iS[\fields]}		\nn\\
  &= \int D\fields'\, \fOp^{(1)'}(x_1)\dots \fOp^{(N)'}(x_N) e^{iS[\fields']}		\nn\\
  &= \int D\fields\, \fOp^{(1)}(x_1)\dots \fOp^{(N)}(x_N) e^{iS[\fields]}		\nn\\
  &=  \langle \fOp^{(1)}(x_1)\dots \fOp^{(N)}(x_N) \rangle\ ,
  \label{eq: global WTI, no instantons}
\end{align}
where viewing $\fOp[\fields]$ as a composite function of the fields $\fields$, we have $\fOp'=\fOp[\fields']$. This is the global \ward{} identity for the symmetry $g$. We can equivalently write the infinitesimal form,
\begin{align}
  \sum_{a=1}^N \left\langle \fOp^{(1)}(x_1)\dots \delta_{G}\fOp^{(a)}(x_a)\dots \fOp^{(N)}(x_N) \right\rangle = 0\ ,
  \label{eq: usual ward identities}
\end{align}
for each $G\in \frak{su}(1,3)$.\\

Let us now consider what changes when we allow for instanton insertions. The configuration space of the theory is now the disjoint union of subspaces on which we specify instanton insertions $\{(x_a,n_a)\}$. Hence, in calculating the correlation function of a set of operators $\fOp_a$, we must also specify which of these subspaces we perform the path integral over. Further, within each of these subspaces we encounter a number of zero modes, undamped by the path integral. The bosonic zero modes correspond simply to the space of gauge-inequivalent instantonic gauge field configurations as discussed above, while we also generically expect fermionic zero modes in each such background. We should therefore also specify gauge field and fermionic boundary conditions in a neighbourhood of each $x_a$.

 This leads us to define
\begin{align}
  \left\langle \fOp^{(1)}(x_1)\dots \fOp^{(N)}(x_N) \right\rangle_{\{(x_a,n_a),\, q_a\}} := \int_{\{(x_a,n_a),\, q_a\}} D\fields\, \fOp^{(1)}(x_1)\dots \fOp^{(N)}(x_N) e^{iS[\fields]}\ ,
  \label{eq: generalised path integral}
\end{align}
Here, the path integral is performed only over configurations $\fields$ with instanton insertions $\{(x_a,n_a)\}_{a=1}^N$. We additionally include formal multi-indices $q_a$ which specify asymptotic field behaviour in a neighbourhood of the $x_a$, corresponding to the bosonic and fermionic instanton moduli as mentioned above, and about which we will have more to say in Section \ref{subsec: instanton operators}. Note, the operator insertion points are the same as the instanton insertion points, denoted $x_a$. This is done without loss of generality, since we allow for any of the operators $\fOp^{(a)}$ to be the identity operator $\mathds{1}$, and we allow any of the $n_a$ to vanish.

Next, consider some $SU(1,3)$ transformation $g$, with corresponding transformed fields $\fields'(x) = g\fields(x)$. If $\fields$ has instanton insertions $\{(x_a,n_a)\}$, then by (\ref{eq: moving of instanton insertions}) we have that $g\fields$ has instanton insertions $\{(x_ag,n_a)\}$. It is important to note that this then induces a right group action of $SU(1,3)$ on the boundary data $q_a$. In particular, if the fields $\fields$ have instanton insertions $\{(x_a,n_a)\}$ with boundary data $q_a$, we can define $q_a g$ as the boundary data of $g\fields$ near $x_a g$. Hence, again assuming no non-trivial Jacobian factor, we  have
\begin{align}
  \int_{\{( x_ag^{-1}, n_a),\, q_a g^{-1}\}} D\fields = \int_{\{(x_a,n_a),\, q_a\}} D\fields'\ .
  \label{eq: change of integration variable}
\end{align}
Then, consider in particular $g$ lying in the subgroup of $SU(1,3)$ generated by $\{P_-, P_i, B, C^\alpha, T\}$, for which we additionally have $S[\fields']=S[\fields]$. We have then
\begin{align}
  \langle \fOp^{(1)'}(x_1)\dots \fOp^{(N)'}(x_N) \rangle_{\{( x_a g^{-1}, n_a),\, q_a g^{-1}\}} &= \int_{\{(x_ag^{-1}, n_a),\, q_a g^{-1}\}} D\fields\, \fOp^{(1)'}(x_1)\dots \fOp^{(N)'}(x_N) e^{iS[\fields]}		\nn\\
  &= \int_{\{(x_a,n_a),\, q_a\}} D\fields'\, \fOp^{(1)'}(x_1)\dots \fOp^{(N)'}(x_N) e^{iS[\fields]}		\nn\\
  &= \int_{\{(x_a,n_a),\, q_a\}} D\fields'\, \fOp^{(1)'}(x_1)\dots \fOp^{(N)'}(x_N) e^{iS[\fields']}		\nn\\
  &= \int_{\{(x_a,n_a),\, q_a\}} D\fields\, \fOp^{(1)}(x_1)\dots \fOp^{(N)}(x_N) e^{iS[\fields]}		\nn\\
  &= \langle \fOp^{(1)}(x_1)\dots \fOp^{(N)}(x_N) \rangle_{\{(x_a, n_a),q_a\}}\ .
  \label{eq: WT identity nice generators}
\end{align}
which is a generalisation of (\ref{eq: global WTI, no instantons}).

We now consider the rest of $SU(1,3)$. The only difference here is that we no longer necessarily have $S[\fields']=S[\fields]$. First consider $g=\exp\left( \epsilon^i M_{i+} \right)$. Then, we have
\begin{align}
  \langle \fOp^{(1)'}(x_1)\dots \fOp^{(N)'}(x_N) \rangle_{\{( x_a g^{-1}, n_a),\, q_a g^{-1}\}}& \nn\\
  &\hspace{-20mm}= \int_{\{(x_ag^{-1}, n_a),\, q_a g^{-1}\}} D\fields\, \fOp^{(1)'}(x_1)\dots \fOp^{(N)'}(x_N) e^{iS[\fields]}		\nn\\
  &\hspace{-20mm}= \int_{\{(x_a,n_a),\, q_a\}} D\fields'\, \fOp^{(1)'}(x_1)\dots \fOp^{(N)'}(x_N) e^{iS[\fields]}		\nn\\
  &\hspace{-20mm}= \prod_{a=1}^N \left( \frac{\overline{\mathcal{M}_{-\epsilon}(x_a)}}{\mathcal{M}_{-\epsilon}(x_a)} \right)^{kn_a}\langle \fOp^{(1)}(x_1)\dots \fOp^{(N)}(x_N) \rangle_{\{(x_a, n_a),\, q_a\}}\ .
  \label{eq: WT identity M}
\end{align}
Following the same steps, for $g=\exp\left( \epsilon K_+ \right)$ we have
\begin{align}
  \langle \fOp^{(1)'}(x_1)\dots \fOp^{(N)'}(x_N) \rangle_{\{( x_a g^{-1}, n_a),\, q_a g^{-1}\}}& \nn\\
  &\hspace{-20mm} = \prod_{a=1}^N \left( \frac{1+2\epsilon \bar{z}(x_a,0)}{1+2\epsilon z(x_a,0)} \right)^{kn_a}\langle \fOp^{(1)}(x_1)\dots \fOp^{(N)}(x_N) \rangle_{\{(x_a, n_a),\, q_a\}}\ .
  \label{eq: WT identity K}
\end{align}
Hence, through (\ref{eq: WT identity M}) and (\ref{eq: WT identity K}) we find that in the quantum theory, we still have global \ward{} identities corresponding to $M_{i+}, K_+$. But these identities are deformed from the naive result (\ref{eq: global WTI, no instantons}), which holds only in an absence of instanton insertions.

\subsection{An Alternative Perspective, and Instanton Operators}\label{subsec: instanton operators}

Before moving on to find the more general local counterparts to these \ward{} identities, let us describe an equivalent but nonetheless useful notation we may use to denote instanton insertions in the quantum theory. This reformulation, in terms of \textit{instanton operators}, will in particular allow for a compact infinitesimal form of (\ref{eq: WT identity nice generators})--(\ref{eq: WT identity K}), while also making contact with previous work in Lorentzian Yang-Mills theories in five dimensions \cite{Lambert:2014jna,Tachikawa:2015mha,Bergman:2016avc}.\\

In the previous Section, we chose to introduce the notion of instanton insertions in the quantum theory by specifying the path integration domain. At least at a formal level, we could instead have expanded the space of operators in the theory. Let us denote by $\Phi_n(x)$ a local operator which, in addition to carrying some representation of $SU(1,3)$, also carries charge under the $U(1)$ topological current $\text{tr}(F\wedge F)$. In detail, such an operator satisfies
\begin{align}
  \left\langle \left(\frac{1}{8\pi^2}\int_{S^4(x)}\text{tr}(F\wedge F)\right)\fOp_n(x) \dots  \right \rangle  = \left\langle n\fOp_n(x)\dots\right\rangle
\end{align}
for some $n\in \mathbb{Z}$, where here $S^4(x)$ is a 4-sphere surrounding $x$, sufficiently small such that it does not enclose any other insertions.

However, if we are now to reproduce the path integral manipulations that lead to the \ward{} identities (\ref{eq: WT identity nice generators})---(\ref{eq: WT identity K}), we need some prescription for how such topologically charged operators are constructed in practise in terms of the fields of the theory.

In analogy with monopole operators appearing in three-dimensional gauge theory \cite{Borokhov:2002ib}, one can construct the operators $\fOp_n$ through the introduction of disorder operators known as instanton operators \cite{Bergman:2016avc,Tachikawa:2015mha,Lambert:2014jna}. Then, the inclusion in the path integral of some instanton operators $\I_n^{\{q\}}(x)$ is defined in terms of our previous notation by
\begin{align}
  \int D\fields\,\, \I_{n_1}^{\{q_1\}}(x_1)\dots \I_{n_N}^{\{q_N\}}(x_N)\Big( \dots \Big) = \int_{\{(x_a,n_a),\,q_a\}}\Big( \dots \Big)
\end{align}
In particular, this path integral vanishes identically unless $\sum_a n_a=0$. 

It is natural at this point to say a little more about the formal index $q$, and in particular its interpretation in canonical quantisation. In some quantisation of the theory, $\I_n^{\{q\}}$ is the creation operator of an instanton-particle. Precisely what state is created is specified by the index $q$. In a pure gauge theory, this index would correspond to the physical (as opposed to gauge-redundant) collective coordinates of an $n$-instanton on $S^4$ in $SU(N_c)$, which although complicated are accessible by virtue of the ADHM construction \cite{Atiyah:1978ri}. However, in a theory with fermions such as the theory considered here, we generically have fermion zero modes in an instanton background, giving rise to a degenerate ground state. Thus, in acting with $\I_n^{\{q\}}$ on the vacuum, we need the index $q$ to specify which of these vacuum states is created. The full classification of these fermion zero modes, and thus a precise formulation of the index $q$, has been achieved for the case of a single $SU(N_c)$ instanton \cite{Tachikawa:2015mha}, providing $\I_{\pm 1}^{\{q\}}$. A more general treatment remains an important open problem in five-dimensional gauge theory. For the purposes of this paper, it is sufficient to assume the existence of such a complete formulation.\\

The $SU(1,3)$ transformation properties of $\I_n^{\{	q\}}$ are then induced by that of the spacetime and boundary data $q$, with $\I_n^{\{q\}'}(x)=g\I_n^{\{q\}}(x) := \I^{\{qg^{-1}\}}_n(xg^{-1})$. The change in path integral measure (\ref{eq: change of integration variable}) is hence recast simply as 
\begin{align}
  \int D\fields\,\, \I^{\{q_1\}'}_{n_1}(x_1)\I^{\{q_2\}'}_{n_2}(x_2)\dots \I^{\{q_N\}'}_{n_N}(x_N)\Big( \dots \Big) = \int D\fields'\,\, \I_{n_1}(x_1)\I_{n_2}(x_2)\dots \I_{n_N}(x_N)\Big( \dots \Big)\ .
  \label{eq: I path integral transformation}
\end{align}
With such a formulation in place, we can now build an operator carrying instanton charge, as
\begin{align}
  \fOp_n(x) = \I_n^{\{q\}}(x) \fOp(x)
\end{align}
where $\fOp=\fOp[\fields]$ is once again simply some composite function of the fields $\fields$ and their derivatives.

Then, $\fOp_n(x)$ transforms in a representation of $SU(1,3)$, which is a tensor product of the representations of $\I_n^{\{q\}}$ and $\fOp(x)$. Given some $g\in SU(1,3)$, we have $\fOp_n'(x)=g\fOp_n(x) = \I_n^{\{qg^{-1}\}}(xg^{-1}) (g\fOp)(x)=\fOp_n(x) + \epsilon \delta_G \fOp_n(x)$ where $g=e^{\epsilon G}$. In particular, $\delta_G \fOp_n(x)$ as always takes the form $\delta_G \fOp_n(x) = - G_\partial \fOp_n(x) - r_G(x) \fOp_n(x)$ for differential operator $G_\partial$ and matrix $r_G(x)$, and defines a representation of $\frak{su}(1,3)$. Note, we define, for instance, $\partial_i \I^{\{q\}}_n(x)$ by requiring $\langle \partial_i \I^{\{q\}}_n \rangle = \partial_i\langle  \I^{\{q\}}_n \rangle $. Further, as with fields, for the sake of later notations convenience we trivially define $\delta_{P_+}\I_n^{\{q\}}(x)=0$, so that $\fOp_n$ sits in a representation of $\frak{h}=\frak{u}(1)\oplus \frak{su}(1,3)$ in which $P_+$ is trivially represented.

We can then reproduce each of the \ward{} identity derivations of the previous Section, with for instance the manipulation from the first to second line of (\ref{eq: WT identity nice generators}) being now of the form (\ref{eq: I path integral transformation}). We thus arrive at simply
\begin{align}
  \langle \fOp_{n_1}^{(1)'}(x_1)\dots \fOp_{n_N}^{(N)'}(x_N) \rangle =  \langle \fOp_{n_1}^{(1)}(x_1)\dots \fOp_{n_N}^{(N)}(x_N) \rangle\ ,
  \label{eq: global WTI, no instantons, new notation}
\end{align}
when $g$ lies in the subgroup of $SU(1,3)$ generated by $\{P_-, P_i, B, C^\alpha, T\}$, or infinitesimally, 
\begin{align}
  \sum_{a=1}^N \left\langle \delta_{G} \fOp_{n_a}^{(a)}(x_a) \prod_{b\neq a} \fOp_{n_b}^{(b)}(x_b) \right\rangle = 0\ .
  \label{eq: infinitesimal undeformed WTIs}
\end{align}
For transformations generated by the remaining two generators $G=M_{i+},K_+$ we have
\begin{align}
  \langle \fOp_{n_1}^{(1)'}(x_1)\dots \fOp_{n_N}^{(N)'}(x_N) \rangle= \prod_{a=1}^N \left( \frac{\overline{\mathcal{M}_{-\epsilon}(x_a)}}{\mathcal{M}_{-\epsilon}(x_a)} \right)^{kn_a}\langle \fOp_{n_1}^{(1)}(x_1)\dots \fOp_{n_N}^{(N)}(x_N) \rangle\ .
  \label{eq: global M WTI, new notation}
\end{align}
and
\begin{align}
  \langle \fOp_{n_1}^{(1)'}(x_1)\dots \fOp_{n_N}^{(N)'}(x_N) \rangle  = \prod_{a=1}^N \left( \frac{1+2\epsilon \bar{z}(x_a,0)}{1+2\epsilon z(x_a,0)} \right)^{kn_a}\langle \fOp_{n_1}^{(1)}(x_1)\dots \fOp_{n_N}^{(N)}(x_N) \rangle\ .
  \label{eq: global K WTI, new notation}
\end{align}
respectively. These then have the infinitessimal forms
\begin{align}
  \sum_{a=1}^N \left\langle \Big( \delta_{M_{i+}} + ikn_a x_a^i  \Big)\big( \fOp_{n_a}^{(a)}(x_a) \big)\prod_{b\neq a} \fOp_{n_b}^{(b)}(x_b) \right\rangle &= 0		\ ,\nn\\
  \sum_{a=1}^N \left\langle  \Big( \delta_{K_+} + ikn_a x_a^i x_a^i \Big)\big( \fOp_{n_a}^{(a)}(x_a) \big)\prod_{b\neq a} \fOp_{n_b}^{(b)}(x_b) \right\rangle &= 0		\ .
  \label{eq: infinitesimal deformed WTIs}
\end{align}

\subsection{Local \ward{} Identities}\label{subsec: local WTIs}

Having now seen that symmetry is restored in the quantum theory, in which \ward{} identities are deformed in the presence of instanton operators, let us now present the much more general \textit{local} \ward{} identities. These will in particular determine the corresponding Noether currents.

We derive the identities following the standard procedure. We consider the variation of correlation functions under a broader class of transformations, in which the $\su(1,3)$ variations are allowed to vary locally according to some function $\epsilon(x)$. Note however that $\epsilon(x)$ must still be approximately constant in a neighbourhood of the points $x_a$, to ensure that the resulting transformations still map into the extended configuration space. Then, taking the functional derivative with respect to $\epsilon(x)$ of the resulting expression, for each $G\in\su(1,3)$ we arrive at
\begin{align}
  &-i\,\left\langle \mathcal{W}_G(x) \prod_{a=1}^N  \fOp_{n_a}^{(a)}(x_a) \right\rangle = \star\sum_{a=1}^N \delta^{(5)}(x-x_a) \left\langle \delta_G\fOp_{n_a}^{(a)}(x_a) \prod_{b\neq a}\fOp_{n_b}^{(b)}(x_b) \right\rangle\ ,
  \label{eq: general local WTI}
\end{align}
Note, we have for simplicity restricted to operators $\fOp_{n_a}^{(a)}$ that depend only on the fields $A, X^I, \Psi$ and not their derivatives. More generally, one would find additional terms one the right-hand side of the form $\partial \left( \text{contact term} \right)$.

For $G\in\{P_-, P_i, B, C^\alpha, T\}$, the top forms $\mathcal{W}_G$ are given by
\begin{align}
  \mathcal{W}_G =  d \star J_G\ ,
\end{align}
for Noether currents $J_G$. Once again, the story is different for $M_{i+}, K_+$, for which we find
\begin{align}
  \mathcal{W}_{M_{i+}} 	&= d \star J_{M_{i+}} + x^i\, d\left( \frac{k}{8\pi^2}  \text{tr}\left( F\wedge F \right) \right)		\nn\\
  						&= d \star J_{M_{i+}} + k\star \sum_{a=1}^N n_a x_a^i \delta^{(5)}(x-x_a)\ ,
\end{align}
and
\begin{align}
  \mathcal{W}_{K_+} 	&= d \star J_{K_+} + x^i x^i\, d\left( \frac{k}{8\pi^2}  \text{tr}\left( F\wedge F \right) \right)		\nn\\
  						&= d \star J_{K_+} + k\star \sum_{a=1}^N n_a x_a^i x_a^i \delta^{(5)}(x-x_a)\ .
\end{align}
The explicit forms of the Noether currents $J_G$ can be found in Appendix \ref{app: Noether currents}.

It is natural then to reorganise terms in (\ref{eq: general local WTI}) for $G=M_{i+}, K_+$, to bring the set of \ward{} identities to a more familiar form. We have
\begin{align}
  &-i\,d\star\left\langle J_G(x) \prod_{a=1}^N \fOp_{n_a}^{(a)}(x_a) \right\rangle= \star\sum_{a=1}^N \delta^{(5)}(x-x_a) \left\langle \tilde{\delta}_G\fOp_{n_a}^{(a)}(x_a) \prod_{b\neq a}\fOp_{n_b}^{(b)}(x_b) \right\rangle\ ,
  \label{eq: general local WTI with tildes}
\end{align}
where the new variations $\tilde{\delta}_G$ act as
\begin{align}
  \tilde{\delta}_G \fOp_n(x) 			&=		\delta_G\fOp_n(x)\quad \text{for } G\in\{P_-, P_i, B, C^\alpha, T\}		\ ,\nn\\
  \tilde{\delta}_{M_{i+}} \fOp_n(x)	&= \delta_{M_{i+}} \fOp_n(x) + iknx^i \fOp_n(x)									\ ,\nn\\
  \tilde{\delta}_{K_+} \fOp_n(x)	&= \delta_{K_+} \fOp_n(x) + iknx^ix^i \fOp_n(x)	\ ,
  \label{eq: tilde variations of I}
\end{align}
Then, by integrating (\ref{eq: general local WTI with tildes}) over $\mathbb{R}^5$ and taking suitable boundary conditions on $S^4_\infty$, we recover the global \ward{} identities (\ref{eq: infinitesimal undeformed WTIs}) and (\ref{eq: infinitesimal deformed WTIs}), written compactly in terms of the $\tilde{\delta}_G$ as
\begin{align}
  \sum_{a=1}^N \left\langle \tilde{\delta}_{G}\fOp_{n_a}^{(a)}(x_a) \prod_{b\neq a} \fOp_{n_b}^{(b)}(x_b) \right\rangle = 0\ .
  \label{eq: infinitesimal WTIs with tilde}
\end{align}
Let us summarise our findings so far. The classical theory admitted an $SU(1,3)$ spacetime symmetry in the absence of instanton insertions. The corresponding infinitesimal variation of fields is denoted $\delta_G$ for each $G\in\su(1,3)$, which form a representation of $\su(1,3)$ when acting on the gauge field $A$ and matter fields $X^I, \Psi$. We extended this to include a variation $\delta_{P_+}$ that acts trivially on all fields $\delta_{P_+}\fields=0$, and in this way realised the $\{\delta_G\}$ as a representation of $\frak{h}$, with brackets  $[\delta_{G_1},\delta_{G_2}]=\delta_{[G_1,G_2]}$ as in (\ref{eq: extended su(1,3) algebra})--(\ref{eq: rotation subalgebra action}).

We found that this symmetry was broken in the classical theory in the presence of instanton operators. However, this breaking is local to the instanton insertion points $x_a$, and thus the resulting \ward{} identities in the quantum theory could nonetheless be written in the standard form (\ref{eq: general local WTI with tildes}) in terms of Noether currents $J_G$. Integrating these local identities over $\mathbb{R}^5$, we recovered the infinitesimal form of the global \ward{} identities (\ref{eq: infinitesimal undeformed WTIs})--(\ref{eq: infinitesimal deformed WTIs}).

The \ward{} identities (\ref{eq: general local WTI with tildes}) are written not in terms of our original variations $\delta_G$, but instead in terms of variations $\tilde{\delta}_G$, which we have defined for each $G\in \mathcal{B}=\{P_-, P_i, B, C^\alpha, T, M_{i+}, K_+\}$. In particular, they differ from the $\delta_G$ for $G=M_{i+}, K_+$ when acting on operators carrying non-zero instanton charge, as in (\ref{eq: tilde variations of I}). 

We are then lead to ask: are the $\{\tilde{\delta}_G\}_{G\in \mathcal{B}}$ the generators of a representation of $\frak{su}(1,3)$ under commutation, like the $\delta_G$ are? The answer is in fact \textit{no}. In particular, we find a \textit{single} commutator that does not close on $\frak{su}(1,3)$, which is
\begin{align}
  [\tilde{\delta}_{M_{i+}},\tilde{\delta}_{P_j}] \fOp_n(x)  &= \left( -\tfrac{1}{2}\Omega_{ij} \tilde{\delta}_T - 2\delta_{ij} \tilde{\delta}_B+\Omega_{ik} \eta^I_{jk} \tilde{\delta}_{C^\alpha} - ik\, \delta_{ij}  n \right)\fOp_n(x)	\ .
  \label{eq: deformed tilde bracket}
\end{align}
Suppose however that we define a new variation $\tilde{\delta}_{P_+}$ that acts as
\begin{align}
  \tilde{\delta}_{P_+}\fOp_n(x) = ikn\,\fOp_n(x)\ .
\end{align}
Equivalently, we have that no fields in the theory are charged under $\tilde{\delta}_{P_+}$, but instanton operators transform as $\tilde{\delta}_{P_+}\I_n^{\{q\}}(x) = ikn\, \I_n^{\{q\}}(x)$. Then, we have
\begin{align}
  [\tilde{\delta}_{M_{i+}},\tilde{\delta}_{P_j}]\fOp_n(x) &= \left( -\delta_{ij} \tilde{\delta}_{P_+}-\tfrac{1}{2}\Omega_{ij} \tilde{\delta}_T - 2\delta_{ij} \tilde{\delta}_B+\Omega_{ik} \eta^I_{jk} \tilde{\delta}_{C^\alpha}\right)\fOp_n(x)	\ .
\end{align}
Then, by direct comparison with the algebra (\ref{eq: extended su(1,3) algebra}), we find quite remarkably that the full set of variations $\{\tilde{\delta}_G\}_{\mathcal{B}\cup \{P_+\}}$ \textit{do} generate a representation of $\frak{h}$, with the operator $\fOp_n$ now carrying charge $ikn$ under $\tilde{\delta}_{P_+}$. In other words, the operator $\fOp_n$ carries Kaluza-Klein momentum in an emergent sixth dimension.

We can then organise our space of operators into primaries and descendants of $\frak{h}$ \cite{Lambert:2020zdc}. In particular, if $\fOp_n$ is a primary operator then we have
\begin{align}
  	\tilde{\delta}_{P_+}\left(\I_n\fOp\right)		\ &=	 \ ikn \left(\I_n\fOp\right)		\ ,												\nn\\
  	\tilde{\delta}_{P_-}\left(\I_n\fOp\right)		\ &=	 \ -\left( P_- \right)_\partial \left(\I_n\fOp\right)	\ ,							\nn\\
  	\tilde{\delta}_{P_i}\left(\I_n\fOp\right)		\ &= \ -\left( P_i \right)_\partial \left(\I_n\fOp\right)	\ ,							\nn\\
  	\tilde{\delta}_{B}\left(\I_n\fOp\right)		\ &=	 \ -\left( B \right)_\partial \left(\I_n\fOp\right) - r_\fOp[B]   \left(\I_n\fOp\right)\ ,	\nn\\
  	\tilde{\delta}_{C^\alpha}\left(\I_n\fOp\right)	\ &=	 \ -\left( C^\alpha \right)_\partial \left(\I_n\fOp\right) - r_\fOp[C^\alpha] \left(\I_n\fOp\right)	\ ,\nn\\
  	\tilde{\delta}_{T}\left(\I_n\fOp\right)		\ &= \	-\left( T \right)_\partial \left(\I_n\fOp\right) -  \Delta \left(\I_n\fOp\right)	\ ,\nn\\
  	\tilde{\delta}_{M_{i+}}\left(\I_n\fOp\right)	\ &= \	-\left( M_{i+} \right)_\partial \left(\I_n\fOp\right) - \left(\tfrac{1}{2}\Delta \Omega_{ij} x^j - ikn x^i	 +2x^i r_\fOp[B] - \Omega_{ik} \eta^\alpha_{jk} x^j r_\fOp[C^\alpha]\right) \left(\I_n\fOp\right)			\ ,\nn\\
  	\tilde{\delta}_{K_+}\left(\I_n\fOp\right)			\ &=	 \ -\left( K_+ \right)_\partial \left(\I_n\fOp\right)  - \left(2\Delta\, x^- - ikn x^i x^i +2x^i x^i r_\fOp[B] - x^i x^j \Omega_{ik}\eta^\alpha_{jk} r_\fOp[C^\alpha]   \right) \left(\I_n\fOp\right) \ .
  	\label{eq: delta tilde on modes}
\end{align}
for scaling dimension $\Delta$ and spin $\{r_\fOp[B],r_\fOp[C^\alpha]\}$.

We can now extend the local \ward{} identity to read once again
\begin{align}
  &-i\,d\star\left\langle J_G(x) \prod_{a=1}^N \fOp_{n_a}^{(a)}(x_a) \right\rangle = \star\sum_{a=1}^N \delta^{(5)}(x-x_a) \left\langle \tilde{\delta}_G \fOp_{n_a}^{(a)}(x_a)\prod_{b\neq a}\fOp_{n_b}^{(b)}(x_b) \right\rangle\ ,
  \label{eq: final local WTIs}
\end{align}
which now holds for all $G\in\mathcal{B}\cup \{P_+\}$, where we define
\begin{align}
  J_{P_+} = -\frac{k}{8\pi^2}\star \tr\left( F\wedge F \right)\ .
  \label{eq: P_+ Noether current}
\end{align}
It is indeed straightforward to see that for $G=P_+$, (\ref{eq: final local WTIs}) is satisfied trivially. Further, (\ref{eq: final local WTIs}) holds for all $G\in\frak{h}$ with $J_{G_1+G_2}=J_{G_1}+J_{G_2}$ and $J_{[G_1,G_2]} = \tilde{\delta}_{G_1} J_{G_2} - \tilde{\delta}_{G_2} J_{G_1}$ for all $G_1,G_2\in\frak{h}$. Integrating over $\mathbb{R}^5$, we once again arrive at the global identities
\begin{align}
  \sum_{a=1}^N \left\langle \tilde{\delta}_{G}\fOp_{n_a}^{(a)}(x_a)\prod_{b\neq a} \fOp_{n_b}^{(b)}(x_b) \right\rangle = 0\ ,
  \label{eq: final global infinitesimal WTIs}
\end{align}
which hold for all $G\in\frak{h}$. We can equivalently write this in its finite form, as
\begin{align}
 	 \big\langle \fOp_{n_1}^{(1)'}(x_1)\dots \fOp_{n_N}^{(N)'}(x_N) \big\rangle = \left\langle \fOp_{n_1}^{(1)}(x_1)\dots \fOp_{n_N}^{(N)}(x_N) \right\rangle\ ,
 	 \label{eq: final global finite WTIs}
\end{align}
where in this expression, $\fOp_n'(x) = \exp(\tilde{\delta}_G)\fOp_n(x)$. The explicit forms of these finitely-transformed operators can be found in \cite{Mouland:2021urv}. Using these forms, it is in particular straightforward to then see that (\ref{eq: final global finite WTIs}) reproduces the results (\ref{eq: global WTI, no instantons, new notation}), (\ref{eq: global M WTI, new notation}) and (\ref{eq: global K WTI, new notation}).

\subsection{General Solution to \ward{} Identities}

The algebra $\frak{h}$, its representations  and the solutions of the resulting \ward{} identities (\ref{eq: final global infinitesimal WTIs}) have already been studied extensively \cite{Lambert:2020zdc}. Thus we can readily apply those results here. For instance, if we consider a pair of scalar operators $\fOp_{n_1}^{(1)},\fOp_{n_2}^{(2)}$ of the theory with scaling dimensions $\Delta_1,\Delta_2$ and instanton charge $n_1,n_2$, respectively, the resulting 2-point function is fixed up to an overall constant. It is given by
\begin{align}
  \big\langle \fOp_{n_1}^{(1)}(x_1) \fOp_{n_2}^{(2)}(x_2) \big\rangle = \delta_{\Delta_1,\Delta_2}\delta_{0,n_1+n_2}d(\Delta_1,n_1)\frac{1}{(z_{12}\bar{z}_{12})^{\Delta_1/2}}\left(\frac{z_{12}}{\bar{z}_{12}}\right)^{n_1}\ ,
  \label{eq: 2pt WTI solution}
\end{align}
for some constant $d(\Delta_1,n_1)$, and $z_{12}=z(x_1, x_2)= -\bar{z}_{21}$ as defined in (\ref{zis}).

One can then continue to find the general solution at $N$-points. This takes a form familiar from regular conformal field theory: a pre-factor which solves the inhomogeneous \ward{} identities, multiplied by an undetermined function $H$ of $\frak{su}(1,3)$ invariant  combinations of coordinates. Explicitly, we have \cite{Lambert:2020zdc}
\begin{align}
	&\langle \fOp_{n_1}^{(1)}(x_1) \dots \fOp_{n_N}^{(N)}(x_N) \rangle \ \nn \\
	&\qquad= \ \delta_{0,n_1+\dots+n_N}  \, \left[ \prod_{a<b}^N (z_{ab} \bar{z}_{ab})^{-\alpha_{ab}/2}  \left( \frac{z_{ab}}{\bar{z}_{ab}} \right)^{(n_a-n_b)/N}\right]  H\left( \frac{|z_{ab}||z_{cd}|}{|z_{ac}||z_{bd}|}, \frac{z_{ab}z_{bc}z_{ca}}{\bar{z}_{ab}\bar{z}_{bc}\bar{z}_{ca}}  \right) \ .
	\label{eq: general N-point function}
\end{align}
where $z_{ab}=z(x_a, x_b)$. The constants $\alpha_{ab}=\alpha_{ba}$ satisfy $\sum_{b\neq a}\alpha_{ab} = \Delta_a$ for each $a=1,\dots, N$, and for a suitable choice of the function $H$ can be taken to be
\begin{align}
  \alpha_{ab} = \frac{1}{N-2}(\Delta_a+ \Delta_b) - \frac{1}{(N-1)(N-2)}\sum_{a=1}^N \Delta_a
\end{align}
for all $N\ge 3$.

The full set\footnote{A proof that this list is exhaustive can be found in \cite{Mouland:2021urv}.} of $\frak{su}(1,3)$-invariant objects fall into two categories; the familiar cross-ratios $|z_{ab}||z_{cd}|/|z_{ac}||z_{bd}|$ of which there are $N(N-3)/2$, and the more novel phases \linebreak $z_{ab}z_{bc}z_{ca}/\bar{z}_{ab}\bar{z}_{bc}\bar{z}_{ca}$, of which there are $(N-1)(N-2)/2$. In particular, even at $N=3$ there is a single invariant phase, and thus in contrast to regular conformal field theory, the 3-point function is fixed only up to a function of one variable.

\subsection{The Quantisation of $k$}\label{sec: k}

A necessary requirement that the Lagrangian (\ref{eq: Lagrangian}) gives rise to a well-defined quantum field theory is that $e^{i S[\fields]}$ is a single-valued functional on the theory's configuration space. Such a constraint can have deep and subtle implications, especially in a theory with non-trivial topological sectors. Take for instance the three-dimensional Abelian Chern-Simons theory, whose action $S_\text{CS}$ is not gauge invariant in the presence of monopole fluxes, and thus fails to be single-valued on configuration space---defined to be the space of fields modulo gauge transformations. Nonetheless, $e^{iS_\text{CS}}$ remains single-valued even in the presence of monopole fluxes, \textit{provided} that the Chern-Simons level is quantised in the integers.

In this Section, we will prove a comparable result for the theory defined by Lagrangian (\ref{eq: Lagrangian}). In detail, we will prove the following claim:
\begin{center}
\begin{minipage}{0.79\textwidth}
	Take the configuration space to be a union over sectors of arbitrary instanton insertions. Then, a necessary condition such that $e^{i S[\fields]}$ is single-valued on this configuration space is that $k\in \frac{1}{2}\mathbb{Z}$.
\end{minipage}
\end{center}
Our proof is constructive, and in particular does not provide a complete picture of the global properties of the action as a functional on configuration space. 

Let us outline the steps taken to demonstrate the claim. We will first define a one-parameter family of fields configurations $\fields_\gamma$, $\gamma\in\mathbb{R}$. In particular, we demonstrate explicitly that this one-parameter family of configurations in fact defines a closed loop in configuration space; in other words, it satisfies $\fields_{\gamma+2\pi} = \fields_\gamma$, and so in particular $\fields_{2\pi}=\fields_0$.

Next, we will compute an explicit expression for $\exp \left(i S[\fields_\gamma]\right)$ as a function of $\exp \left(i S[\fields_0]\right)$. Using this, we find examples of closed loops such that 
\begin{align}
  \exp \left(i S[\fields_{2\pi}]\right) = e^{4\pi k i}\exp \left(i S[\fields_0]\right)
  \label{eq: e^{iS} phase shift}
\end{align}
Thus, $e^{iS[\fields]}$ is generically a multi-valued functional on configuration space. Note, the construction of a loop satisfying (\ref{eq: e^{iS} phase shift}) will necessarily require that the configurations $\fields_\gamma$ have non-trivial instanton insertions, and thus this phase ambiguity only arises when we allow for such non-trivial topological sectors.
Hence, we find that for this particular loop in configuration space, $e^{iS[\fields]}$ is single-valued only for $k\in \frac{1}{2}\mathbb{Z}$, thus proving the claim.\\

So let us now explicitly construct the closed loop in configuration space. Note that we have already computed the finite variation of the action under $SU(1,3)$ transformations, in particular finding rather suggestive forms (\ref{eq: finite action variation M}) and (\ref{eq: finite action variation K}) for the transformations under $M_{i+}$ and $K_+$, respectively. As such, we can utilise these results---and thus simplify our calculations here---by seeking a closed loop in configuration space that lies within the $SU(1,3)$ orbits.

Let us then consider the following one-parameter family of $SU(1,3)$ elements,
\begin{align}
  h(\gamma) = \exp\left[\gamma\left(\frac{1}{2}K_+ + P_-\right)\right]\ .
\end{align}
Making use of the expressions in Appendix B of \cite{Lambert:2019fne}, one can show that if we take the $\frak{su}(1,3)$ generators to lie in the fundamental representation, then $h(\gamma+2\pi)=h(\gamma)$. Thus, $h(\gamma)$ defines closed loop in the fundamental representation of $SU(1,3)$, of period $2\pi$.

Next we want to consider the finite transformation of coordinates and fields under the $SU(1,3)$ transformation $h(\gamma)$. Once again, we can make our lives easier by utilising known results. First, we can use the Baker-Campbell-Hausdorff formula to show that for all $\gamma\in (-\pi/2,\pi/2)$,
\begin{align}
  h(\gamma) &= \exp\Big[ \tfrac{1}{2}\left(\tan\gamma\right) K_+ \Big] \exp\Big[\left(\sin\gamma \cos\gamma\right) P_- \Big] \exp\Big[\log\left(\sec\gamma\right) T \Big]		\nn\\
  &= \exp\Big[\log\left(\cos\gamma\right) T \Big]	 \exp\Big[ \left(\sin\gamma \cos\gamma\right) P_- \Big] \exp\Big[ \tfrac{1}{2}\left(\tan\gamma\right) K_+ \Big]\ .
  \label{eq: h factors}
\end{align}
Thus, we can compute the finite variation of coordinates and fields under $h(\gamma)$ by performing successive finite transformations under elements generated purely by $T, P_-$ and $K_+$.

First let us consider the coordinates, for which the finite transformations under $e^{\epsilon T},e^{\epsilon P_-}$ and $e^{\epsilon K_+}$ can be found in Appendix A of \cite{Mouland:2021urv}. Then, we find
\begin{align}
  \left(xh^{-1}(\gamma)\right)^- 		&= \frac{\cos 2\gamma\, x^- - \frac{1}{2}\sin 2\gamma\left(1-\left(\left(x^-\right)^2 + \frac{1}{16}|\vex|^4\right)\right)}{\left(\cos\gamma+ \sin\gamma \,x^-\right)^2+\frac{1}{16}\sin^2\gamma\, |\vex|^4}				\nn\\
  \left(xh^{-1}(\gamma)\right)^i		&= 	\frac{ \cos\gamma\, x^i + \frac{1}{4}\sin\gamma \left(4x^- x^i - |\vex|^2 \Omega_{ij} x^j\right)}{\left(\cos\gamma+\sin\gamma \,x^-\right)^2+\frac{1}{16} \sin^2\gamma\, |\vex|^4}\ ,
  \label{eq: xh}
\end{align}
which are valid for all $\gamma\in\mathbb{R}$. Note then that we do indeed have $xh^{-1}(\gamma+2\pi) = xh^{-1}(\gamma)$, and so in particular $xh^{-1}(2\pi) = x$.

Let us now consider the orbits in configuration space generated by $h(\gamma)$. We take some starting configuration $\fields = \{A, X^I, \Psi, G_{ij}\}$, and then consider the new configuration $h(\gamma)\fields$ obtained by transforming by the $SU(1,3)$ element $h(\gamma)$. Explicitly,
\begin{align}
  h(\gamma)\fields = \{h(\gamma)A_-, h(\gamma)A_i, h(\gamma)X^I, h(\gamma)\Psi, h(\gamma)G_{ij}\}\ .
\end{align}
The form of $h(\gamma)\fields$ is then found by exponentiating the known infinitesimal variation of each field under $(\frac{1}{2}\delta_{K_+}+ \delta_{P_-})$, as given in Appendix \ref{app: field variations}. However, we should be cautious: it is a priori not clear that we have $h(\gamma+2\pi)\fields = h(\gamma)\fields$, as for instance fields may lie in a projective representation of $SU(1,3)$.

It is straightforward to see that this is not the case for the gauge field $A=(A_-, A_i)$ and scalars $X^I$. For the gauge field, we can simply write down the form of the transformed fields, which take the standard form
\begin{align}
  \left(h(\gamma)A_-\right)(x) &= \left[\partial_-\! \left(xh^{-1}(\gamma)\right)^-\right] A_-\left(xh^{-1}(\gamma)\right) + \left[\partial_-\! \left(xh^{-1}(\gamma)\right)^i\right] A_i\left(xh^{-1}(\gamma)\right)		\nn\\
  \left(h(\gamma)A_i\right)(x)  &= \left[\partial_i\! \left(xh^{-1}(\gamma)\right)^-\right] A_-\left(xh^{-1}(\gamma)\right) + \left[\partial_i\! \left(xh^{-1}(\gamma)\right)^j\right] A_j\left(xh^{-1}(\gamma)\right)	\ .
\end{align}
Then  the $2\pi$ periodicity of $xh(\gamma)^{-1}$ as seen in (\ref{eq: xh}) ensures that we do indeed have $h(\gamma+2\pi)A = h(\gamma)A$. Since our task is to exhibit a closed path in configuration space under which the action is not invariant it is enough to set the scalars and Fermions to zero, since if $X^I,\Psi=0$ then $h(\gamma)X^I, h(\gamma)\Psi=0$ for all $\gamma$.

Finally, we need to address the Lagrange multiplier field $G_{ij}$. As previously mentioned, the infinitesimal variation of $G_{ij}$ under the generators of $SU(1,3)$ is somewhat subtle, and in particular we do not know a closed form for its finite variation. However, we can work around this issue in the following way. Our ultimate aim is to compute $S[h(\gamma)\fields]$ as a function on $S[\fields]$. Now recall, the Lagrangian (\ref{eq: Lagrangian}) depends on $G_{ij}$ only through the term $G_{ij} \mathcal{F}_{ij} = G_{ij} \mathcal{F}^+_{ij}$, with $\mathcal{F}^+_{ij} = \frac{1}{2}\left(\mathcal{F}_{ij} + \frac{1}{2}\varepsilon_{ijkl} \mathcal{F}_{kl} \right)$. So let us suppose that the gauge field $A$ of our starting configuration $\fields$ satisfies $\mathcal{F}^+_{ij}=0$, and thus $S[\fields]$ is independent of $G_{ij}$. Then, crucially, the constraint $\mathcal{F}^+_{ij}=0$ is an $SU(1,3)$ invariant: for all $g\in SU(1,3)$, if $A$ satisfies $\mathcal{F}^+_{ij}=0$ then so does the transformed field $gA$. Hence, we have that for all $\gamma$, $S[h(\gamma)\fields]$ is independent of $G_{ij}$.

Thus, making contact with the language of the above proof outline, let us define the starting configuration $\fields_0 = \{A, X^I=0, \Psi=0, G_{ij}\}$, while the orbit in configuration space is defined to be
\begin{align}
  \fields_\gamma = \{h(\gamma)A, X^I=0,  \Psi=0, G_{ij}\}\ .
\end{align}
In particular, we do not transform $G_{ij}$. Then, we have already shown that $\fields_{\gamma+2\pi} = \fields_\gamma$ as desired.

Finally, we are ready to compute $S[\fields_\gamma]$. First note that, following the discussion above, the fact that the configuration $\fields$ satisfies the constraint $\mathcal{F}^+_{ij}=0$ ensures that we have $S[\fields_\gamma] = S[h(\gamma) \fields]$. We can then leverage the factorisation (\ref{eq: h factors}) along with the finite $K_+$ transformation (\ref{eq: finite action variation K}) to simply write down
\begin{align}
  \exp \big(i S[\fields_\gamma]\big) = e^{i S[\fields]} \prod_{a=1}^N  u_\gamma(x_a;n_a)\ ,
\end{align}
where 
\begin{align}
  u_\gamma(x;n) = \left(\frac{\cos\gamma - \sin\gamma\,\bar{z}(x,0)}{\cos\gamma - \sin\gamma\,z(x,0)}\right)^{kn}\ ,
\end{align}
and the configuration $\fields_\gamma$ has instanton insertions $\{(x_a,n_a)\}_{a=1}^N$. We remind the reader that $z(x_1,x_2)$ is defined in (\ref{zis}).

We can now ask what happens as we pass from $\gamma=0$ through to $\gamma=2\pi$. Then, for all $x=(x^-, \vex)$ with $\vex\neq \vec{0}$, we find that as we pass from $\gamma=0$ to $\gamma=2\pi$, the combination $(\cos\gamma - \sin\gamma\,\bar{z}(x,0))$ encircles the origin of the complex plane clockwise precisely once. Thus, we have
\begin{align}
  u_{\gamma+2\pi} (x;n) = e^{4\pi k n i} u_{\gamma} (x;n)\ .
\end{align}
Conversely, if $\vex=\vec{0}$, then $u_\gamma(x;n) = 1$ for all $\gamma$, and thus trivially $u_{\gamma+ 2\pi}(x;n)= u_\gamma(x;n)$.\\

Suppose first then that all instanton insertions $\{(x_a,n_a)\}$ of the starting configuration $\fields$ lie away from the spatial origin, {\it i.e.}\ $\vex_a\neq \vec{0}$ for all $a=1,\dots, N$. Then, we have
\begin{align}
   \exp \big(i S[\fields_{2\pi}]\big) = \exp \left(4\pi k i \sum_{a=1}^N n_a\right) e^{i S[\fields]} = e^{i S[\fields]}\ ,
\end{align}
 since $\sum_a n_a=0$. Something special happens, however, if any of the instanton insertions lie at the spatial origin. 
In particular, taking a single insertion at the origin with charge $-1$, we find
\begin{align}
   \exp \big(i S[\fields_{2\pi}]\big) = e^{4\pi k i}  e^{i S[\fields]}\ ,
   \label{eq: anomalous variation of e^{iS}}
\end{align}
as promised in the outline above. More generally, if the sum of the charges of all instanton insertions at the origin is $m\in\mathbb{Z}$, the relevant phase factor is $e^{-4\pi k m i}$, and thus the above case is in this sense minimal. Thus, from (\ref{eq: anomalous variation of e^{iS}}) we find that a necessary condition such that $e^{iS[\fields]}$ is single-valued on configuration space is that $k$ takes values in the half-integers, $k\in \frac{1}{2}\mathbb{Z}$.

Next in Section \ref{sec: 6D} we will require $k\in\mathbb{Z}$ in order that the theory admits a six-dimensional interpretation. It is unclear whether any such novel interpretation holds when $k\in \{\frac{1}{2},\frac{3}{2},\dots\}$ or whether a more refined argument to the one above, perhaps by including fermions, can be made to exclude these cases.

\section{Reconstructing Six Dimensions}\label{sec: 6D}

We have seen that the five-dimensional path integral based on the Lagrangian (\ref{eq: Lagrangian}) leads to a theory with an $SU(1,3)\times U(1)$ symmetry that acts non-trivially on a  Kaluza-Klein-like tower of operators obtained by inserting instantons.   Furthermore the associated \ward{} identities are naturally solved by the  Fourier modes of a conformally compactified six-dimensional conformal field theory with the role of Kaluza-Klein momentum replace by instanton number. Thus we now would like to reconstruct the correlators of a six-dimensional theory from the five-dimensional path integral.

Owing to the $2\pi$ interval over which $x^+$ runs, such an interpretation requires the eigenvalues of $P_+$ to take discrete integer values \cite{Lambert:2020zdc}. This is indeed the case, so long as $k\in\mathbb{Z}$. Then, $\tilde{\delta}_{P_+}\fOp_n(x) =ikn\,\fOp_n(x)$, which precisely identifies $\I_n(x)\fOp(x)$ as the $(kn)^\text{th}$ Fourier mode of some six-dimensional operator. In particular, a choice of $k=  1$ allows for the realisation of the full spectrum of Fourier modes on the conformal compactification, while higher $k$ corresponds to a $\mathbb{Z}_{k}$ orbifold thereof.

We can now form a coherent state of Fourier modes, and so define the notion of a six-dimensional operator in our theory. We can then ask when such operators can be interpreted as those of a six-dimensional conformal field theory.

\subsection{Constructing Six-dimensional Observables}\label{sec: Construction}

Given a collection of local operators $\{\fOp_n(x)\}_{n\in \mathbb{Z}}$, we are lead to define six-dimensional operator 
\begin{align}
  \sOp(x^+, x^-, x^i):= \sum_{n\in\mathbb{Z}} e^{-iknx^+} \fOp_n(x^-, x^i)\ ,
  \label{eq: six d operators definition}
\end{align}
for some new coordinate $x^+$. Then, we have 
\begin{align}
  \tilde{\delta}_{P_+} \sOp (x^+, x^-, x^i) = -\frac{\partial}{\partial x^+}\sOp (x^+, x^-, x^i)\ ,
\end{align}
and so $P_+$ is identified as translations along an emergent sixth dimension\footnote{Note, the sign here is consistent with the convention used to define the $\tilde{\delta}_G$, since for instance we had $\tilde{\delta}_{P_-}\sOp(x)=-\partial_-\sOp(x)$.}. 

Indeed, it is straightforward to go a step further, and show that for generic $G\in\frak{h}$, we have $\tilde{\delta}_G\sOp(x) = - G _\partial^\text{6d} \sOp(x) - r^\text{6d}(x)\sOp(x)$, where as usual $r^\text{6d}$ acts on any indices of $\sOp$, while the six-dimensional vector fields $G_\partial^\text{6d}$ form precisely the algebra of conformal Killing vector fields of six-dimensional Minkowski space which commute with $(P_+)^\text{6d}_\partial=\partial_+$. Explicitly, these are
\begin{align}
	\left( P_+ \right)^\text{6d}_\partial \ 	&= \ \partial_+  \, ,  \nn\\
	\left( P_- \right)^\text{6d}_\partial \ 	&= \ \partial_-  \, ,  \nn\\
	\left( P_i \right)^\text{6d}_\partial \ 	&= \ \tfrac{1}{2}\Omega_{ij} x^j \partial_- + \partial_i	\, , \nn\\
	\left( B \right)^\text{6d}_\partial 		\ &= \ -\tfrac{1}{2}\,\Omega_{ij}x^i\partial_j 		\, , \nn\\
	\left( C^I \right)^\text{6d}_\partial 	\ &= \ \tfrac{1}{2}\eta^I_{ij}x^i\partial_j 								\, , \nn\\
	\left( T \right)^\text{6d}_\partial 		\ &= \ 2x^- \partial_- + x^i \partial_i					\, , \nn\\
	\left( M_{i+} \right)^\text{6d}_\partial 	\ &= \ x^i \partial_+ + \left( \tfrac{1}{2}\Omega_{ij} x^- x^j - \tfrac{1}{8} x^j x^j x^i \right)\partial_- + x^- \partial_i  +\tfrac{1}{4}( 2\Omega_{ik}x^k x^j + 2\Omega_{jk}x^k x^i - \Omega_{ij}x^k x^k )\partial_j	\, , \nn\\
	\left( K_{+} \right)^\text{6d}_\partial 	 \ &= \ x^i x^i \partial_+ + ( 2 ( x^- )^2 - \tfrac{1}{8}   ( x^i x^i )^2 )\partial_- +( \tfrac{1}{2} \Omega_{ij} x^j x^k x^k + 2 x^- x^i )\partial_i	\, .
	\label{eq: 6d algebra vector field rep}
\end{align}
as first derived in \cite{Lambert:2019fne}.\\

Let us finally review how one can reconstruct operators defined on six-dimensional space with the standard Minkowski metric \cite{Lambert:2020zdc}. First, one must perform a Weyl rescaling  in order to arrive at operators $\hat{\sOp}$,
\begin{align}
\hat{\sOp}( x) &= \cos^{\hat\Delta}(x^+/2)\sOp(x)	\nonumber\\
&= \cos^{\hat\Delta}(x^+/2)\sum_{n\in\mathbb{Z}} e^{-iknx^+}\fOp_n(x^-, x^i)\ ,\label{resumC}
\end{align}
where $\hat \Delta$ is the six-dimensional scaling dimension\footnote{Note, in general the operator $\hat{\sOp}$ constructed this way would not have definite eigenvalue under the six-dimensional dilatation. However, this can be guaranteed by ensuring that the $SU(1,3)$ representations of the $\fOp_n$ differ only by their charge under the central element $P_+$. }.
The prefactor corresponds to the non-trivial conformal factor relating the metric (\ref{eq: metric}) to the Minkowski metric $ds^2_\text{M}$, as $ds^2 = \cos^2(x^+/2)ds^2_\text{M}$. 

One may then want to perform a coordinate transformation to standard coordinates $\hat x$ of six-dimensional Minkowski space, such that $ds^2_\text{M}=-2d\hat{x}^+d\hat{x}^-+d\hat{x}^i d\hat{x}^i$. These are related to the $(x^+, x^-, x^i)$ by
\begin{align}
x^+ &= 2\arctan\left(\frac{\hat{x}^+}{2}\right)	\ ,\nonumber\\
x^- &= \hat{x}^- - \frac{\hat{x}^+\hat{x}^i \hat{x}^i}{2\left(4+ (\hat{x}^+)^2\right)}\ ,\nonumber\\
x^i &= \frac{4\left(\hat{x}^i +\frac12\Omega_{ij}\hat{x}^j \hat{x}^+\right)}{4+(\hat x^+)^2}\ .
\end{align}
Let us consider for example a scalar operator $\hat{\sOp}$ with six-dimensional scaling dimension $\hat{\Delta}$, for which this coordinate transformation is trivial. Following (\ref{eq: six d operators definition}), this should be built from modes $ \fOp_n$ that are scalars of $\frak{h}$, and which have Lifshitz scaling dimension $\Delta=\hat{\Delta}$. Then, we can write\footnote{The binomial expansion from the second to third lines only holds for $\Delta\in \mathbb{Z}$.}
\begin{align}
\hat{\sOp}(\hat x) &= \cos^{\Delta}(x^+ /2)\sOp(x )	\nn\\
&= 2^{-\Delta}\left(e^{ix^+/2}+e^{-ix^+/2}\right)^\Delta\sum_{n\in\mathbb{Z}} e^{-iknx^+} \fOp_n(x^- , x^i )\nonumber\\
&= 2^{-\Delta}\sum_{n\in\mathbb{Z}}\sum_{l=0}^\Delta{\Delta \choose l} e^{i\Delta x^+/2-i(kn+l)x^+} \fOp_n(x^- , x^i )\nonumber\\
&= 2^{-\Delta}\sum_{n\in\mathbb{Z}}\sum_{l=0}^\Delta {\Delta \choose l}  \left(\frac{1+i\hat x^+}{\sqrt{1+(\hat x^+)^2}}\right)^{\Delta  - 2(kn+l)} \fOp_n(x^-(\hat x) , x^i(\hat x) )\ .
\end{align}
Thus we have a way to construct six-dimensional operators out of five-dimensional ones. Furthermore in principle we can compute their  correlation functions using the five-dimensional path integral {\it viz}: 
 \begin{align}
 \langle\hat{\sOp}^{(1)}(\hat x_1)\ldots\hat{\sOp}^{(N)}(\hat x_N)\rangle = 	\cos^{\Delta_1}(x^+_1 /2)\ldots\cos^{\Delta_N}(x^+_N /2)
  \sum_{n_1,\ldots,n_N\in\mathbb{Z}} e^{-ik(n_1x^+_1+\ldots+n_Nx^+_N)}\nonumber\\
  \times\langle\fOp_{n_1}^{(1)}(x^-_1 , x^i_1 )\ldots \fOp_{n_N}^{(N)}(x^-_N , x^i_N)\rangle\ .\end{align}
again for six-dimensional scalars $\hat{\sOp}^{(a)}$ of scaling dimension $\hat{\Delta}_a = \Delta_a$. A generalisation to higher spin operators is conceptually straightforward, but we do not explore it here.

\subsection{Topological \ward{} Identities and the Full Conformal Algebra} \label{fullconf}

We have learnt that for $k\in\mathbb{Z}$, the theory (\ref{eq: Lagrangian}) is able to describe Fourier modes on the $x^+$ interval of momentum $kn$ for any $n\in\mathbb{Z}$. Such a mode $\fOp_n$ can in turn be in principle constructed by dressing a local operator of the theory with an instanton operator of charge $n$. Thus, for general $k\in\mathbb{Z}$, we may propose that the theory is a six-dimensional conformal field theory on the orbifold $\mathbb{R}^{1,5}/\mathbb{Z}_k$, where the $\mathbb{Z}_k$ quotient acts on the $x^+$ interval as $x^+\to x^++2\pi/k$. This leads to a rather curious orbifold in terms of the familiar six-dimensional coordinates $\hat x$. In general, such an orbifold breaks the full conformal algebra $\frak{so}(2,6)$ precisely down to $\frak{h}=\frak{su}(1,3)\oplus \frak{u}(1)$. We have  shown that this is indeed the symmetry obeyed by the theory.

Then  something special must happen at  $k=1$, where there is no orbifold. In particular, in this case the full six-dimensional conformal symmetry is not broken, and so the theory should realise the full $\frak{so}(2,6)$. We learn that a necessary condition for the theory (\ref{eq: Lagrangian}) to describe a six-dimensional conformal field theory is that its symmetries are further enhanced at strong coupling ($k=1$).

Let us briefly note that this enhancement is in some sense analagous with the ABJM theory for M2-branes, where it is the R-symmetry (rather than the spacetime symmetry) that is enhanced from $\frak{su}(4)\oplus \frak{u}(1)$ to $\frak{so}(8)$ at strong coupling. Further details of this analogy can be found in \cite{Lambert:2020zdc,Mouland:2021urv}.\\

So let us fix $k=1$. Our aim now is to explore how this symmetry enhancement should happen. Once again, we will study the theory through the lens of correlation functions, and in particular the partial differential equations they satisfy. But first, let us make some comments on the use of five-dimensional operators $\fOp_n$, which fall into representations of $\frak{h}\subset \frak{so}(2,6)$, as building blocks in the realisation of $\frak{so}(2,6)$ representations.

One can consider the action of $\frak{so}(2,6)$ on a six-dimensional operator $\hat{\sOp}$, which is easily translated to an action on the Weyl rescaled $\sOp$. This operator is then in turn decomposed as
\begin{align}
  \sOp(x^+, x^-, x^i) = \sum_{n\in\mathbb{Z}} e^{-inx^+} \fOp_{n}(x^-, x^i)
  \label{eq: generic Fourier expansion}
\end{align}
for Fourier modes $\fOp_n$ satisfying $\tilde{\delta}_{P_+} \fOp_n = in \fOp_n$. The fact that the Fourier expansion breaks $\frak{so}(2,6)\to \frak{h}=\frak{su}(1,3)\oplus \frak{u}(1)$ is precisely the statement that it is only the subalgebra $\frak{h}$ of infinitesimal transformations of $\sOp$ that act on each Fourier level independently; recall, $\frak{h}$ is simply the maximal subalgebra that commutes with $P_+$. This is in contrast to the rest of $\frak{so}(2,6)$, which generically scramble up the Fourier modes. In other words, the $\fOp_n$ fall into representations of $\frak{h}$. As previously mentioned, when $\hat{\sOp}$ is an $\frak{so}(2,6)$ scalar primary of scaling dimension $\hat{\Delta}$, one can show that under $\frak{h}$, the $\fOp_n$ must transform precisely as scalar primaries as defined with respect to $\frak{h}$ ({\it i.e.}\ $r[B]=0, r[C^\alpha]=0$), with Lifshitz scaling dimension $\Delta$ under $T$ given simply by the original six-dimensional scaling dimension, $\Delta=\hat{\Delta}$, and of course $P_+$ momentum $n$.

We then turn to the question of how to identify the modes $\fOp_n$ of a given operator $\sOp$ we wish to reconstruct. The first task here is to deduce the required $\frak{h}$ representation of $\fOp_n$ given the six-dimensional quantum numbers of $\sOp$. This is conceptually straightforward, and in some cases practically immediate too; for instance, a scalar primary in six-dimensions is built of five-dimensional scalar primaries of $\frak{h}$.

With this done, we must now look at the space of operators of the form $\fOp_n = \I_n^{\{q\}}\fOp$, where $\fOp$ is some composite of the fields $\fields$, and classify them by their $\frak{h}$ representations. To do so at general $n$ is an open problem, which in a supersymmetric theory such as ours amounts to determining which supermultiplet the instanton operator $ \I_n^{\{q\}}$ lies in. As previously mentioned, this has been solved in Lorentzian $SU(N_c)$ theories only for $n=\pm 1$ \cite{Tachikawa:2015mha}, while no results currently exist for non-Lorentzian theories such as ours. Thus, tackling this issue is an important next step in our programme.

Supposing that we have successfully identified a class of five-dimensional operators in the correct $\frak{h}$ representations, we would now like to derive some criteria by which we can correctly organise them into a particular six-dimensional local operator. Thus, in the rest of this Section  we will derive further conditions that must be satisfied by the correlators of the $\fOp_n$, which we expect to be crucial in identifying them precisely.

Note, our focus for the remainder of this Section is on the reconstruction of \textit{local} operators in six dimensions. It is clear however that one can also in principle reconstruct extended six-dimensional operators, including those extended along the $x^+$ interval. Such constructions---which will in particular be essential to the construction of defect operators in six-dimensional CFTs---are left to future work. \\

So suppose that we think we have correctly identified in our theory the Fourier modes $\fOp^{(a)}_n$ of some six-dimensional scalar primaries $\hat{\sOp}^{(a)}$ of scaling dimension $\hat{\Delta}_a$ under the six-dimensional dilatation. These $\fOp^{(a)}_n$ are then necessarily scalars of $\frak{h}$ of Lifshitz dimension $\Delta_a=\hat{\Delta}_a$, and so in turn will generically be some linear combination of dimension $\Delta_a$ scalar primaries of the form $\I_n^{\{q\}}\fOp$. We now want some purely five-dimensional criteria by which we can check whether we've chosen the $\fOp^{(a)}_n$ correctly. We know that the correlation functions $\langle \hat{\sOp}^{(1)} \hat{\sOp}^{(2)}\dots \hat{\sOp}^{(N)} \rangle$ satisfy the \ward{} identities of $\frak{so}(2,6)$, while a priori we have only so far shown that the five-dimensional correlators $\langle \fOp^{(1)}_{n_1}\fOp^{(2)}_{n_2}\dots \fOp^{(N)}_{n_N} \rangle$ satisfy those corresponding to the subalgebra $\frak{h}$. The additional criteria that the $\fOp_n^{(a)}$ must satisfy is that the six-dimensional correlators they resum to satisfy the full set of $\frak{so}(2,6)$ identities. Let's see how this works in practise.

Suppose then we take the six-dimensional \ward{} identity for some $G\in\frak{so}(2,6)$ that lies outside $\frak{h}$. It is then conceptually straightforward to expand the $\hat{\sOp}^{(a)}$ in terms of the $\fOp^{(a)}_n$, and thus determine the corresponding equations satisfied by the correlators of the $\fOp^{(a)}_n$. However, in contrast to identities (\ref{eq: infinitesimal WTIs with tilde}) corresponding to elements of $\frak{h}$, this new identity will necessarily be a partial differential equation relating correlators with \textit{different} Fourier mode numbers. From the perspective of the five-dimensional theory, these are non-trivial relations between correlators calculated in distinct topological sectors, and thus we refer to such identities as \textit{topological \ward{} identities} (TWTIs). 

It is instructive to look at a particular example of some $G\in \frak{so}(2,6)$ that lies outside $\frak{h}$. We consider the $\Npt$-point function $\langle \hat{\sOp}^{(1)} \hat{\sOp}^{(2)}\dots \hat{\sOp}^{(N)} \rangle$. We know that this $\Npt$-point function satisfies the \ward{} identities of $\frak{so}(2,6)$, spanned by $\{P_\mu^\sixd, M_{\mu\nu}^\sixd, D^\sixd, K_\mu^\sixd\}$, with $\mu\in\{+,-,i\}$. Let us focus on the six-dimensional dilatation $D^\sixd$, which is not preserved by the Fourier decomposition\footnote{The five-dimensional Lifshitz scaling $T\in\frak{h}$ is found as $T= D^\sixd - M^\sixd_{+-}$.}; in other words, $D^\sixd\notin \frak{h}$. Our aim is to understand how the invariance of the six-dimensional $N$-point function under $D^\sixd$ manifests in the correlators of the $\fOp^{(a)}_n$.

Let us now denote by $\delta_D \hat\sOp^{(a)}$ the infinitesimal variation of $\hat\sOp^{(a)}$ under $D^\sixd$. Explicitly, we have
\begin{align}
  \tilde{\delta}_D \hat{\Op}^{(a)}(\hat{x}) = -D_\partial \hat{\Op}^{(a)}(\hat{x}) - \Delta_a \hat{\Op}^{(a)}(\hat{x})
\end{align}
in terms of the vector field
\begin{align}
  D_\partial 	&= \hat{x}^+ \hat{\partial}_+ + \hat{x}^- \hat{\partial}_- + \hat{x}^i \hat{\partial}_i		\nn\\
  				&= \sin (x^+) \partial_+ + \left(x^--\tfrac{1}{4}\sin(x^+) |\vec x|^2\right)\partial_- + \tfrac{1}{2}\big( \left(1+\cos(x^+)\right)x^i + \sin(x^+) \Omega_{ij} x^j \big)\partial_i \ .
\end{align}
The variation of the Fourier modes $\fOp^{(a)}_n$ is then fixed by 
\begin{align}
  \delta_D\hat{\sOp}^{(a)} = \cos^{\Delta_a}\! \left(\tfrac{x^+}{2}\right) \sum_{n\in\mathbb{Z}} e^{-inx^+} \delta_D\fOp^{(a)}_n\ .
\end{align}
Explicitly, we find
\begin{align}
  \delta_D \fOp^{(a)}_n = -\tfrac{1}{2}\tilde{\delta}_T\fOp^{(a)}_n - \mathcal{D}_-\fOp^{(a)}_{n-1} - \mathcal{D}_+\fOp^{(a)}_{n+1}\ ,
  \label{eq: delta D On}
\end{align}
where we define
\begin{align}
  \mathcal{D}_+\fOp^{(a)}_n &= \left(-\frac{1}{2}n+\frac{1}{4}\Delta_a + \frac{i}{8}|\vec x|^2 \partial_- + \frac{1}{4}x^i \partial_i - \frac{i}{4}\Omega_{ij} x^j \partial_i\right)\fOp^{(a)}_n		\nn\\
  \mathcal{D}_-\fOp^{(a)}_n &= \left(+\frac{1}{2}n+\frac{1}{4}\Delta_a - \frac{i}{8}|\vec x|^2 \partial_- + \frac{1}{4}x^i \partial_i + \frac{i}{4}\Omega_{ij} x^j \partial_i\right)\fOp^{(a)}_n\ ,
\end{align}
while we can read off $\tilde{\delta}_T \fOp^{(a)}_n=\delta_T \fOp^{(a)}_n$ from (\ref{eq: delta tilde on modes}), as
\begin{align}
  \tilde{\delta}_T \fOp_n^{(a)} = - (T)_\partial \fOp_n^{(a)} - \Delta_a \fOp_n^{(a)} \ .
\end{align}
As expected, we see that variation under $D^\sixd\notin \frak{h}$ mixes the Fourier modes of $\hat\sOp^{(a)}$.

So now let us consider the \ward{} identity for $D^\sixd$, taking the form \linebreak $\sum_a \langle \hat{\sOp}^{(1)}\dots \delta_D \hat{\sOp}^{(a)} \dots \hat{\sOp}^{(N)}  \rangle = 0 $. Then, expanding in modes and further using the fact that $\langle \fOp^{(1)}_{n_1} \dots \fOp^{(N)}_{n_N} \rangle$ is non-vanishing only for $n_1+\dots + n_N=0$, we find that this \ward{} identity for the $\hat{\sOp}^{(a)}$ is satisfied if and only if we have all of
\begin{align}
  \sum_{a=1}^N \left\langle \fOp^{(1)}_{n_1}(x_1)\dots \tilde{\delta}_T\fOp^{(a)}_{n_a}(x_a) \dots \fOp^{(N)}_{n_N}(x_N) \right\rangle &= 0		\label{eq: T WTI} \ , \\
  \sum_{a=1}^N \left\langle \fOp^{(1)}_{n_1}(x_1)\dots \mathcal{D}_-\fOp^{(a)}_{n_a-1}(x_a)  \dots \fOp^{(N)}_{n_N}(x_N) \right\rangle &= 0		\label{eq: D minus WTI}\ , \\
  \sum_{a=1}^N \left\langle \fOp^{(1)}_{n_1}(x_1)\dots \mathcal{D}_+\fOp^{(a)}_{n_a+1}(x_a)  \dots \fOp^{(N)}_{n_N}(x_N) \right\rangle &= 0	  \label{eq: D plus WTI} \ ,
\end{align} 
for all $n_1,\dots,n_N=1,\dots, N$. Note, the first equation is trivially satisfied as a result of the $P_+$ \ward{} identity whenever $n_1+\dots+n_N\neq 0$, while the following two are similarly trivially satisfied whenever $n_1+\dots+n_N\neq \pm 1$, respectively. 

Now, (\ref{eq: T WTI}) is simply the \ward{} identity for $T\in\frak{h}$, which is satisfied by correlators of the theory by virtue of the symmetries of the action, as we saw in Section \ref{sec: instantons}. In contrast, (\ref{eq: D minus WTI}) and (\ref{eq: D plus WTI}) are new. They are our first explicit examples of TWTIs, as they constitute a non-trivial relationship between correlation functions computed in different topological sectors of the theory. Thus, in order to verify the symmetry enhancement $SU(1,3)\times U(1) \to SO(2,6)$ at strong coupling, one must demonstrate that (\ref{eq: D minus WTI}) and (\ref{eq: D plus WTI}), along with all other TWTIs arising from each of the other elements of $\frak{so}(2,6)$ outside $\frak{h}$, are satisfied.

As we have seen, the two equations (\ref{eq: D minus WTI}) and (\ref{eq: D plus WTI}) descend from the $D^\sixd$ \ward{} identity in six dimensions. It is hopefully evident that one can take identical steps in order to derive further TWTIs that descend from other $G\in\frak{so}(2,6)$ lying outside $\frak{h}$, although we do not explore such other generators in detail here. 

To better illustrate our construction let us   investigate the implications of (\ref{eq: D minus WTI}) and (\ref{eq: D plus WTI}) at 2-points. Recall then, that the functional form of the five-dimensional 2-point function of scalar operators is entirely fixed purely by the WTIs for $\frak{h}$ to be
\begin{align}
  \left\langle \fOp_n(x_1) \fOp_{-n}(x_2) \right\rangle = d(\Delta,n)\frac{1}{(z_{12}\bar{z}_{12})^{\Delta/2}}\left(\frac{z_{12}}{\bar{z}_{12}}\right)^{n}\ .
  \label{eq: 2 point for recurrence}
\end{align}
where the $\fOp_n$ have Lifshitz scaling dimension $\Delta$.
Then, the TWTIs (\ref{eq: D minus WTI}) and (\ref{eq: D plus WTI}) are satisfied precisely if the coefficients $d(\Delta,n)$ satisfy
\begin{align}
  \left(n+\frac{\Delta}{2}\right)d(\Delta,n) - \left(n-\frac{\Delta}{2}+1\right)d(\Delta,n+1)=0\ .
  \label{eq: recurrence relation}
\end{align}
Let us look in particular at the case $\Delta\in 2\mathbb{Z}$.  We then find the general solution
\begin{align}
  d(\Delta,n) = C_+ { n+\frac{\Delta}{2}-1 \choose n - \frac{\Delta}{2}} + C_- { -n+\frac{\Delta}{2}-1 \choose -n - \frac{\Delta}{2}}\ ,
  \label{eq: recurrence solution protected}
\end{align}
for some constants $C_+, C_-$, where we use the convention
\begin{align}
  {\alpha\choose n} = \begin{cases}
 	\frac{\alpha(\alpha-1)\dots(\alpha-n+1)}{n!}	\qquad& n>0		\\
 	1	\qquad& n=0	\\
 	0	\qquad& n<0
 \end{cases}\ .
\end{align}
Thus, by imposing the TWTI for $D^\sixd$, we have determined all of the coefficients $d(\Delta,n)$ up to the two free variables $C_+, C_-$. 

Note however that the six-dimensional 2-point function to which these five-dimensional correlators must resum depends on only a single overall normalisation.   However, it turns out that (\ref{eq: recurrence solution protected}) provides the general solution to the full set of TWTIs. To understand this, note that the additional degree of freedom arises because the five-dimensional 2-point functions (\ref{eq: 2 point for recurrence}) are the Fourier modes of a \textit{Lorentzian} six-dimensional 2-point function. Crucially, such a 2-point function admits two different values---often parameterised by a suitable $i\epsilon$ prescription \cite{Lambert:2020zdc,Hartman:2015lfa}---depending on the ordering of the two operators. Then, the first term in (\ref{eq: recurrence solution protected}) corresponds to one choice of ordering, while the second corresponds to the other.

The important point here is that \ward{} identities and their solutions are blind to such an ordering, and so will produce most generally a linear combination of all possible orderings. Instead, one must rely on the path integral formulation (\ref{eq: generalised path integral}), or else some other quantisation of the theory, to fix the ordering of operators.  In particular if we choose the ordering corresponding to $C_-=0$ then we find
\begin{align}
  \left\langle \fOp_n(x_1) \fOp_{-n}(x_2) \right\rangle =  C_+ { n+\frac{\Delta}{2}-1 \choose n - \frac{\Delta}{2}}\frac{1}{(z_{12}\bar{z}_{12})^{\Delta/2}}\left(\frac{z_{12}}{\bar{z}_{12}}\right)^{n}\ .
  \label{eq: 2 point for recurrence with coefficients}
\end{align}
As a consistency check, we can then use this result to deduce the 2-point function of the six-dimensional operator $\hat{\sOp}=\cos^\Delta(x^+/2)\sOp$, with $\sOp$ written in terms of the $\fOp_n$ as in (\ref{eq: generic Fourier expansion}). As shown in \cite{Lambert:2020zdc}, upon performing the sum over modes explicitly with suitable $i\epsilon$ regularisation, we find
\begin{align}
	 \big\langle \hat{\sOp}(\hat x_1) \hat{\sOp}(\hat x_2) \big\rangle = \frac{(-4)^{-\Delta/2}C_+}{|\hat x_1-\hat x_2|^{2\Delta}}\ .
\end{align}
which is  precisely the correct form for a scalar 2-point function of a six-dimensional conformal field theory.

\section{Conclusions and Future Directions}\label{sec: conclusions}

In this paper we have explored how the path integral based on five-dimensional Lagrangians with an $SU(1,3)$ spacetime symmetry can be used to reconstruct correlation functions of a six-dimensional conformal field theory. In particular we showed how by including non-trivial instanton sectors into the theory the $SU(1,3)$ symmetry of the action is expanded into  $SU(1,3)\times U(1)$ non-perturbatively. We also saw how instanton operators can be used to construct towers of operators which can be identified with suitable Fourier modes of six-dimensional operators. Furthermore we saw that once the instanton sectors are included we must restrict the inverse coupling constant $k\in\tfrac12 {\mathbb Z}$, thereby removing any continuous free parameters. We also explored how imposing the additional symmetries of $SO(2,6)$ that are not present in the five-dimensional theory can be used to constrain the construction of six-dimensional operators. 

While still far from the complete story, we have found an encouraging correspondence between, on one hand, results derived directly from six-dimensional correlators, and on the other hand the allowed topological sectors of the five-dimensional theory (\ref{eq: Lagrangian}). These results support the claim that the path integral formulation (\ref{eq: generalised path integral}) or some refinement thereof will be successful in computing correlators in six-dimensional conformal field theory.

There are many outstanding issues to explore but let us highlight a few. It would be interesting to extend the analysis of Section \ref{fullconf} to other  $SO(2,6)$ \ward{} identities, and then importantly to extend the path integral methods of this paper to demonstrate that this full set is indeed satisfied at strong coupling. Furthermore our theories all enjoy significant supersymmetries which we have not exploited. In particular, the 24 real supercharges realised by the Lagrangian (\ref{eq: Lagrangian}) can be identified with the full set of supercharges in the six-dimensional $(2,0)$ superconformal algebra that are preserved under the $x^+$ reduction \cite{Lambert:2019jwi}, while generalisations with 12 supercharges corresponding to $(1,0)$ superconformal symmetry are also known \cite{Lambert:2020jjm}. Related to this are BPS bounds  and superselection rules. 

In addition it is clearly of interest to compute higher-point functions. In particular 4-point functions are not fixed by conformal symmetry and therefore encode non-trivial six-dimensional dynamics. Computing these should in principle be possible using the path integral methods we have described. Furthermore there has been great progress in the use of localisation techniques to calculate exact results in supersymmetric field theories in a variety of field theories. Our hope is that this can be applied to the Lagrangians discussed here to obtain concrete results for six-dimensional conformal field theories such as the enigmatic $(2,0)$ theory of M5-branes. In so doing we hope to open up a window into the microscopic physics of M-theory.

\section*{Acknowledgements}

R.~Mouland was supported by the STFC studentship ST10837 and David Tong's Simons Investigator grant, A.~Lipstein by the Royal Society as a Royal Society University Research Fellowship holder, and P.~Richmond by STFC grant ST/L000326/1.

\appendix

\section{$\frak{su}(1,3)$ Field Variations}\label{app: field variations}

It is the norm in Lorentzian theories for the action to be written in a manifestly Lorentz-invariant way, with the transformations of fields straightforward to write down. For our theory and its $\frak{su}(1,3)$ spacetime symmetry, we do not have this luxury. The transformations of the fields of the theory can in principle be derived by trial and error. However, there turns out to be an elegant and useful way to derive them from a diffeomorphism-invariant six-dimensional theory, to which the theory is subtly related. The full details of this construction can be found in \cite{Lambert:2019fne}. Here we state the results in a notation more useful for this paper.

We consider a transformation generated by some $G\in\frak{su}(1,3)$. Then, the components of the gauge field transform in a standard way,
\begin{align}
  	\delta_G A_- &= - G_\partial A_- - \big(\partial_- G_\partial^-\big) A_- - \big(\partial_- G_\partial^i\big) A_i	\ ,\nn\\
  	\delta_G A_i &= - G_\partial A_i - \big(\partial_i G_\partial^-\big) A_- - \big(\partial_i G_\partial^j\big) A_j\ ,
\end{align}
{\it i.e.}\ as (minus) the Lie derivative along the vector field $G_\partial$.

The scalar fields $X^I$ also transform under the usual Lie derivative for scalars, except that they are also subject to a compensating Weyl rescaling for $G\in\{T, M_{i+}, K_+\}$. This Weyl factor is given by
\begin{align}
  \omega:= \frac{1}{4}\hat{\partial}_i G_\partial^i\ ,
\end{align}
which takes the values
\begin{align}
  G &= \,\makebox[10mm][l]{$T$}  \longrightarrow \quad \omega = 1		\ ,\nn\\
  G &= \,\makebox[10mm][l]{$M_{i+}$}  \longrightarrow\quad  \omega = \tfrac{1}{2}\Omega_{ij} x^j	\ ,	\nn\\
  G &= \,\makebox[10mm][l]{$K_+$}  \longrightarrow\quad \omega = 2x^-	\ ,
\end{align}
while vanishing for the remaining generators. Then, we have
\begin{align}
  \delta_G X^I = - G_\partial X^I - 2\omega X^I\ .
  \label{eq: X transformation}
\end{align}
This is indeed entirely analogous to the familiar interpretation of usual conformal field theory as a gauge fixing of a theory with both diffeomorphism and Weyl invariance. There, like here, it is a coordinated combination of a diffeomorphism and Weyl rescaling which leaves the metric invariant, and thus forms a symmetry of the gauge fixed theory.

For the fermions, we find
\begin{align}
  \delta_G \Psi = - G_\partial \Psi - \frac{1}{2}\omega\left( 5+\Gamma_{-+} \right) \Psi + \Omega_{ij}( \hat{\partial}_j \omega )\Gamma_+\Gamma_i\Psi + \frac{1}{4}\Lambda^{ij}\Gamma_{ij} \Psi\ ,
\end{align}
where
\begin{align}
  \Lambda^{ij} = (\hat{\partial}_j G_\partial^i) - \omega \delta_{ij} = - \Lambda^{ji}\ .
\end{align}
Explicitly, we find that $\Lambda^{ij}=0$ for $G\in\{P_-, P_i, T\}$, while for the remaining generators,
\begin{align}
  G &= B  &\longrightarrow \quad \Lambda^{ij}&=	\tfrac{1}{2}\Omega_{ij}		\ ,\nn\\
  G &= C^\alpha &\longrightarrow \quad \Lambda^{ij}&=	 - \tfrac{1}{2}\eta^I_{ij}		\ ,\nn\\
  G &= M_{i+} &\longrightarrow \quad \Lambda^{jk}&=	\tfrac{1}{2}\left(\Omega_{jk} x^i+\Omega_{ik}x^j - \Omega_{ij} x^k + \delta_{ik}\Omega_{jl} x^l - \delta_{ij} \Omega_{kl} x^l\right)		\ ,\nn\\
  G &= K_+  &\longrightarrow \quad \Lambda^{ij}&= \tfrac{1}{2}\Omega_{ij} x^k x^k + \Omega_{ik} x^k x^j - \Omega_{jk} x^k x^i		\ .
\end{align}
Then, when acting on the fields $A,X^I, \Psi$, we find that the $\{\delta_G\}$ with\linebreak $G\in \mathcal{B}=\{P_-, P_i, B, C^\alpha, T, M_{i+}, K_+\}$ are precisely the generators of a representation of $\frak{su}(1,3)$.\\

We finally come to the Lagrange multiplier field $G_{ij}$, which arises in a more complicated fashion from the six-dimensional proxy theory. We find
\begin{align}
  \delta_G G_{ij} &= - k^\alpha \partial_\alpha G_{ij} - 4\omega G_{ij} - \left( \Lambda^{mi}G_{mj} - \Lambda^{mj} G_{mi}  \right)			\nn\\
  &\qquad + 2\left( \Omega_{im}(\hat{\partial}_m \omega) F_{-j} - \Omega_{jm}(\hat{\partial}_m \omega) F_{-i}+ \varepsilon_{ijkl} \Omega_{km}(\hat{\partial}_m \omega) F_{-l}  \right)\ ,
\end{align}
We note in particular that, in contrast the other fields, the algebra only closes on $G_{ij}$ on the constraint surface $\mathcal{F}^+=0$. In particular, for each $G_1,G_2\in\mathcal{B}$ we have $[\delta_{G_1},\delta_{G_2}]G_{ij} = \delta_{[G_1,G_2]}G_{ij}$ except for
\begin{align}
  [\delta_{M_{i+}}, \delta_{M_{j+}}]G_{kl} &= \delta_{[M_{i+}, M_{j+}]}G_{kl} + \bar{\delta}_{ij}G_{kl}	\ ,	\nn\\
  [\delta_{M_{i+}}, \delta_{K_+}]G_{jk} &= \delta_{[M_{i+}, K_+]}G_{jk} + 2 x^l\bar{\delta}_{il}G_{jk}\ ,
  \label{eq: Gij algebra extension}
\end{align}
where
\begin{align}
  \bar{\delta}_{ij} G_{kl}			&=2\left(\delta_{ik} \mathcal{F}^+_{jl}-\delta_{il} \mathcal{F}^+_{jk}-\delta_{jk} \mathcal{F}^+_{il}+\delta_{jl} \mathcal{F}^+_{ik}\right)\ .
\end{align}
A discussion of the origin of this extension to the algebra can be found in \cite{Mouland:2021urv}.

Note that we have $\mathcal{F}_{kl}\bar{\delta}_{ij}G_{kl}=0$, which ensures that $\bar{\delta}$ is a symmetry of the Lagrangian. Indeed, since $G_{ij}$ appears only algebraically in $\mathcal{L}$, we have local symmetries $\epsilon(x)\bar{\delta}$ for any function $\epsilon(x)$, and thus we should think of $\bar{\delta}$ as generating an auxiliary gauge symmetry which become trivial on the constraint surface.\\

Finally, note that under Lifshitz scalings as generated by $T$, we have
\begin{align}
  	X^I(x^-, x^i)\quad &\longrightarrow\quad \omega^{-2}  X^I(\omega^{-2}x^-, \omega^{-1}x^i)		\ ,\nn\\
  	A_-(x^-, x^i)\quad &\longrightarrow\quad \omega^{-2}  A_-(\omega^{-2}x^-, \omega^{-1}x^i)		\ ,\nn\\
  	A_i(x^-, x^i)\quad &\longrightarrow\quad \omega^{-1}  X^I(\omega^{-2}x^-, \omega^{-1}x^i)		\ ,\nn\\
  	G_{ij}(x^-, x^i)\quad &\longrightarrow\quad \omega^{-4}  G_{ij}(\omega^{-2}x^-, \omega^{-1}x^i)	\ ,\nn\\
  	\Psi_+(x^-, x^i)\quad &\longrightarrow\quad \omega^{-3}  \Psi_+(\omega^{-2}x^-, \omega^{-1}x^i)	\ ,\nn\\
  	\Psi_-(x^-, x^i)\quad &\longrightarrow\quad \omega^{-2}  \Psi_-(\omega^{-2}x^-, \omega^{-1}x^i)	\ ,
  	\label{eq: Lifshitz scalings of fields}
\end{align}
where we denote by $\Psi_\pm$ the components of $\Psi$ with definite chirality under $\Gamma_{-+}=\Gamma_{05}$, so that $\Gamma_{-+}\Psi_\pm = \pm \Psi_\pm$.

\section{Noether Currents}\label{app: Noether currents}

The Noether currents $J_G$ for $G\in\mathcal{B}$ were first derived in \cite{Lambert:2019fne}, albeit without appreciation for $\delta$-function subtleties due to instanton insertions. We state them here, in a way more consistent with the notation used in this paper. Note that we use $J_G$ to denote a vector field and 1-form interchangeably, as the musical isomorphism with respect to the Euclidean metric on $\mathbb{R}^5$ that relates them is trivial.

 Our expressions are written in terms of the Lagrangian,
\begin{align}
  \mathcal{L} = \frac{k}{4\pi^2} \text{tr} \bigg\{ & \,\frac{1}{2}F_{-i}F_{-i} - \frac{1}{2}\nabla_i X^I \nabla_i X^I + \frac{1}{2} \mathcal{F}_{ij} G_{ij}\nn\\
  &\,\,-\frac{i}{2} \bar{\Psi} \Gamma_+ D_- \Psi + \frac{i}{2} \bar{\Psi} \Gamma_i \nabla_i \Psi - \frac{1}{2}\bar{\Psi} \Gamma_+ \Gamma^I [X^I, \Psi] \bigg\}\ .
\end{align}
For {$G\in\{P_-, P_i, B, C^I, T\}$} we find
\begin{align}
  	J_{G}^- 		&= 	-\left( G_\partial \right)^- \mathcal{L}  +\frac{k}{4\pi^2}\text{tr}\,\bigg[\,\,  -\left( F_{-i} + \tfrac{1}{2}\Omega_{jk} x^k G_{ij} \right)\, \delta_G A_i  -\tfrac{1}{2}\Omega_{ij} x^j \big( \hat{D}_i X^I \big) \delta_G X^I \nonumber\\
  &\hspace{95mm} +\tfrac{i}{2} \bar{\Psi} \left( \Gamma_+ + \tfrac{1}{2} \Omega_{ij} x^j \Gamma_i \right)\delta_G\Psi \,\,\bigg]	\nn\\
  	J_{G}^i 		&= -(G_\partial)^i \mathcal{L} +\frac{k}{4\pi^2}\text{tr}\,\bigg[\,\, \left( F_{-i} + \tfrac{1}{2}\Omega_{jk} x^k G_{ij} \right)\, \delta_G A_-  - G_{ij}\, \delta_G A_j  + \big( \hat{D}_i X^I \big)\, \delta_G X^I \nonumber\\
  &\hspace{122mm} -\tfrac{i}{2}\bar{\Psi} \Gamma_i \delta_G \Psi  \,\,\bigg]\ .		
\end{align}

Whereas for {$M_{i+}$}
\begin{align}
  	J_{M_{i+}}^- 		&= 	\left( -\frac{k}{8\pi^2}x^i \star\text{tr}\left( F\wedge F \right) \right)^-	-\left( M_{i+} \right)_\partial^- \mathcal{L}\nn\\
  	&\qquad  +\frac{k}{4\pi^2}\text{tr}\,\bigg[\,\,  \tfrac{1}{4}x^i X^I X^I-\left( F_{-i} + \tfrac{1}{2}\Omega_{jk} x^k G_{ij} \right)\, \delta_G A_i   \nonumber\\
  &\hspace{45mm}-\tfrac{1}{2}\Omega_{ij} x^j \big( \hat{D}_i X^I \big) \delta_G X^I +\tfrac{i}{2} \bar{\Psi} \left( \Gamma_+ + \tfrac{1}{2} \Omega_{ij} x^j \Gamma_i \right)\delta_G\Psi \,\,\bigg]	\nn\\
  	J_{M_{i+}}^j 		&= 	\left( -\frac{k}{8\pi^2}x^i \star\text{tr}\left( F\wedge F \right) \right)^j	-\left( M_{i+} \right)_\partial^j \mathcal{L}	\nn\\
  	&\qquad+\frac{k}{4\pi^2}\text{tr}\,\bigg[\,\, \tfrac{1}{2}\Omega_{ij} X^I X^I +\left( F_{-i} + \tfrac{1}{2}\Omega_{jk} x^k G_{ij} \right)\, \delta_G A_-   \nonumber\\
  &\hspace{30mm}- G_{ij}\, \delta_G A_j  + \big( \hat{D}_i X^I \big)\, \delta_G X^I -\tfrac{i}{2}\bar{\Psi} \Gamma_i \delta_G \Psi  \,\,\bigg]	\ ,
\end{align}
and {$K_+$}
\begin{align}
  	J_{K_+}^- 		&= 	\left( -\frac{k}{8\pi^2}x^i x^i \star\text{tr}\left( F\wedge F \right) \right)^-	-\left( K_+ \right)_\partial^- \mathcal{L}\nn\\
  	&\qquad  +\frac{k}{4\pi^2}\text{tr}\,\bigg[\,\,  \tfrac{1}{2}x^i x^i X^I X^I-\left( F_{-i} + \tfrac{1}{2}\Omega_{jk} x^k G_{ij} \right)\, \delta_G A_i   \nonumber\\
  &\hspace{45mm}-\tfrac{1}{2}\Omega_{ij} x^j \big( \hat{D}_i X^I \big) \delta_G X^I +\tfrac{i}{2} \bar{\Psi} \left( \Gamma_+ + \tfrac{1}{2} \Omega_{ij} x^j \Gamma_i \right)\delta_G\Psi \,\,\bigg]	\nn\\
  	J_{K_+}^i 		&= 	\left( -\frac{k}{8\pi^2}x^j x^j \star\text{tr}\left( F\wedge F \right) \right)^i	-\left( K_+ \right)_\partial^i \mathcal{L}	\nn\\
  	&\qquad+\frac{k}{4\pi^2}\text{tr}\,\bigg[\,\, \Omega_{ij} x^j X^I X^I +\left( F_{-i} + \tfrac{1}{2}\Omega_{jk} x^k G_{ij} \right)\, \delta_G A_-   \nonumber\\
  &\hspace{30mm}- G_{ij}\, \delta_G A_j  + \big( \hat{D}_i X^I \big)\, \delta_G X^I -\tfrac{i}{2}\bar{\Psi} \Gamma_i \delta_G \Psi  \,\,\bigg]	\ .
\end{align}

\bibliographystyle{JHEP}
\bibliography{2109.04829v2}

\end{document}